%% file: agn_catalog_draft.tex
\documentclass[twocolumn]{aastex631}

\usepackage{colortbl}

\graphicspath{{./}{figures/}}

\usepackage{hyperref}

\begin{document}
\title{The South Pole Telescope AGN Monitoring Campaign:\\ First Release of SPTpol Bright AGN Light Curves}
\input{authors}

\correspondingauthor{John C. Hood II}
\email{hood.astro@gmail.com}

\shortauthors{J.C. Hood II et al.}

\shorttitle{The South Pole Telescope AGN Monitoring Campaign}


\begin{abstract}
The South Pole Telescope (SPT) collaboration has recently embarked upon a campaign to monitor the brightness of a sample of active galactic nuclei (AGN), both in real time and in archival SPT data. The original design of the SPT was optimized for observations of the cosmic microwave background (CMB) at arc-minute and larger angular scales, and it has been used for this purpose for nearly twenty years, using three generations of CMB cameras. Recently it has been recognized that SPT data---and data from other CMB experiments---have the potential to be used for AGN monitoring. In this paper, we present the first public release of data from a full sample of SPT-monitored AGN, comprising 158 AGN light curves and associated data from the SPTpol camera, which was operational from 2012-2016. These light curves were created using observations from the SPTpol 500 deg$^{2}$ survey, in which the instrument was used to scan a 500 deg$^2$ patch of the sky (centered at right ascension $0^\mathrm{h}$, declination $-57.5^\circ$, extending $\pm2$h in R.A.~and $\pm7.5^\circ$ in decl.) several times per day with detectors sensitive to radiation in bands centered at 90 and 150 GHz. We provide a comprehensive description of the observations, the data processing methods, and the resulting light curve catalog. 
As an example of analyses that these data enable, we searched for a correlation between variability and spectral index, and we looked for ``bluer-when-brighter'' trends in the sample. 
Our analysis finds $> 10 \sigma$ significance for correlation between fractional intrinsic variance and mean spectral index in the sample, but no significant evidence for bluer-when-brighter trends. The datasets from this study can be accessed through the SPT Treasury Record of AGN With Historical Activity and Time-Series or \href{https://spt3g.ncsa.illinois.edu/datasets/spt_agn_lightcurves/.}{STRAWHAT} catalog. This initial data release includes light curves derived from SPTpol observations at 90 and 150 GHz, focusing on total intensity. In later updates, SPTpol polarization data and new observations from the SPT-3G instrument at 90, 150, and 220 GHz will be included.
\end{abstract}



\section{Introduction}\label{sec:intro}
Observations of the cosmic microwave background (CMB) provide a cornerstone of our understanding of the universe, and recent and upcoming CMB experiments are measuring the millimeter-wave (mm-wave) sky with ever-increasing sensitivity. Furthermore, these observations are beginning to be used as not just a tool to enrich our understanding of cosmology but also for time-domain astrophysics across multiple CMB experiments, currently including the South Pole Telescope (SPT) and the Atacama Cosmology Telescope (ACT), as discussed in~\citet{guns21, hood23, biermann25, tandoi24}. Among many potential sources for mm-wave time-domain studies, active galactic nuclei (AGN) are of particular interest, since AGN play a crucial role in the process of transforming galaxies and regulating star formation~\citep{ishibashi12,voit15}. They also contribute to the re-ionization and enrichment of the universe with heavy elements~\citep{nyland18}. Current research on AGN includes the categorization of distinct types \citep{urry95}, investigating the origin of AGN jets \citep{blandford19}, and examination of their variable luminosity \citep{agudo11}. However, despite their significance to the structure of our universe, we still have much to learn about the underlying mechanics of these objects. AGN are divided into two main categories of jetted and non-jetted~\citep[e.g.,][]{Padovani17} and many sub-classifications within these classifications~\citep{urry95}. 

To help us understand these sources better, we present a new catalog of AGN light curves that were made with data from the SPT \citep{carlstrom11}. This catalog uses data from the SPT's second-generation camera, SPTpol. This instrument was installed from 2012 to 2016 and was capable of both temperature and polarization measurements of the CMB with observing bands centered at 90 and 150\,GHz~\citep{austermann12}. During its time in use, SPTpol surveyed 500 $\mathrm{deg^2}$ of the southern extragalactic sky at a resolution of arc minutes to milli-Jansky (mJy) level depths~\citep{chou25}. In this work, we choose to monitor the brightest point sources in our survey field, many of which are radio-loud AGN known as ``blazars." Blazars generally have a ratio of radio (5 GHz) to optical (B-band) flux density $\geq$ 10 \citep{kellermann89} and emit strongly at higher energies. These sources are understood to have a relativistic jet pointed at small angles ($<$5 deg) to the observer~\citep{antonucci93, urry95}.

Variability studies serve as a method to better understand the underlying emission processes observed in these sources \citep{Gaur12, hood23, schellenberger24}. The precise origins of gamma-ray emission and flaring events observed across the electromagnetic spectrum, and the underlying physics governing these phenomena, remain undetermined. There are two primary classes of models that seek to explain the origin of gamma-ray emission in blazars, namely the leptonic and hadronic models; these are extensively discussed in~\citet{Boettcher13}. 

In leptonic models, high-energy emission is caused by Compton up-scattering of soft photons by ultra-relativistic electrons that also produce synchrotron emission~\citep{boettcher12, chatterjee12}. Hadronic models, meanwhile, associate the high-energy photons with synchrotron emission by protons, or by secondary particles produced in proton-proton interactions, as discussed in \citet{cerruti20}. If the simplest version of the leptonic hypothesis is correct, gamma-ray flares should always coincide with flares at lower frequencies, in which the synchrotron emission dominates. While this is often seen to be the case, some observations have shown discrepancies, with gamma-ray and lower-frequency flares not always occurring simultaneously \citep{chatterjee13a,chatterjee13b}, indicating the emission processes are more complex than the simplest version of the leptonic model can explain.

There have been many models suggested to explain the multi-wavelength variability in blazars, as discussed in, e.g., \citet{boettcher19}. The following four models included in that reference detail how distinct variability patterns result in AGN light curves. Shock-in-jet (or internal shock) models state that inhomogeneities in the jet flow produce relativistic shocks that travel through the jet plasma and accelerate particles, mostly through diffusive shock acceleration~\citep{marscher85}. Turbulence and magnetic reconnection models suggest that relativistic flows within jets develop turbulence, which may trigger magnetic reconnection and result in distributions of accelerated electrons characterized by hard power-law spectra~\citep{Guo16}. External sources of variability models, in which external matter interacts with the jet, either by direct collision or by means of radiative interactions~\citep{araudo10}. Finally, jet geometries that include bends or helical jets lead to changes in the viewing geometry and the corresponding Doppler factor that manifest as variable emission~\citep{larionov13}.

By supplying additional data, particularly high-cadence multi-wavelength observations across the electromagnetic spectrum (radio to $\gamma$-rays), we can improve our understanding of AGN variability and emission. Such high-cadence observations can be used to create light curves that will help in studies of AGN emission mechanisms and variability at different timescales. In recent years, CMB experiments such as SPT and ACT have begun to provide low noise ($\sim$10-mJy) observations on rapid ($\sim$1-day) cadences over years-long observing campaigns, potentially yielding valuable benefits for source monitoring as a free byproduct of CMB observations. Here, we report the initial results of our AGN monitoring program with SPT, which shows how effectively CMB instruments can be used as tools to provide high-cadence, long-time-series observations of AGN. In this study, we present light curves of 158 bright ($\langle S_{150} \rangle$=38\,mJy,  $\langle S_{90} \rangle$=55\,mJy) AGN.\footnote{These values represent the average flux densities of the entire catalog, where $\langle S_{150} \rangle$ and $\langle S_{90} \rangle$ are the average flux densities of all 158 objects across the two observed frequencies.} The AGN were observed over a period of four years, with minimum signal-to-noise ratio of $\approx$ 1 or greater during a 36-hour time period. 
We refer to this data set as the  SPT Treasury Record of AGN With Historical Activity and Time-Series, or STRAWHAT catalog. Data from the STRAWHAT catalog will enable further investigations of multi-wavelength models of AGN variability by providing mm-wave  measurements from both actively flaring and quiescent states of more than 150 AGN. 

In this paper, we discuss the following details. Section~\ref{sec:obs} describes the SPTpol instrument, survey, and data analysis pipeline that includes our mapping and calibration processes. Section~\ref{sec:sources} details our method for selecting the sources that are included in the STRAWHAT catalog. Section~\ref{sec:buildingLCs} describes our beam characterization, apodization mask, matched filtering, flux density estimation, and light curve processing. Section~\ref{sec:description} describes the catalog itself and data available via the online repository. Section~\ref{sec:discuss} is where we discuss the catalog statistics explored in this study. Additionally, Appendix~\ref{sec:A1} shows a sample of 10 bright AGN light curves, and Appendix~\ref{sec:A2} is a table with basic identifying information for all of the monitored sources.

\section{Instrument, Observations, and Map Making}\label{sec:obs}
In this section, we describe the SPTpol instrument, the 500\,$\mathrm{deg^2}$ survey field, the observing strategy, and the methods used to create the maps and extract the photometric data required to build the STRAWHAT catalog. We describe the map-making and bundling process that takes $\sim$3500 single observations over four years and condenses them into 455 average ``bundle" maps.

\subsection{SPTpol Instrument}\label{sec:inst}
The SPT is dedicated to making low-noise, high-resolution maps of the mm-wave sky, with the primary purpose of mapping the temperature and polarization anisotropies in the CMB \citep{carlstrom11}. The second-generation SPTpol camera was equipped with 180 polarization-sensitive transition-edge-sensor (TES) bolometer pixels sensitive to radiation in a band cenered at roughly 90\,GHz. These detectors were coupled by feedhorn arrays that surrounded 588 polarization-sensitive TES detectors measuring radiation in a band centered at roughly 150\,GHz. The angular resolution of SPTpol was $\sim$1.7 arc minutes at 90\,GHz and $\sim$1.2 arc minutes at 150\,GHz~\citep{austermann12}. The effective band centers for each frequency used in this study were calculated for a flat spectrum ($\alpha$=0). As in, e.g., \citet{reichardt21}, we define the effective band center as the frequency at which the conversion between CMB fluctuation temperature and spectral radiance 
\citep[see, e.g.,][]{planck13-9} is equal to the band-averaged version for a source with a given spectrum.

\subsubsection{\texorpdfstring{SPTpol 500\,$\mathrm{deg^2}$ Survey Observations}{SPTpol 500 deg² Survey Observations}}
\begin{figure*}[htbp]
  \begin{center}
    \includegraphics[width=0.8\textwidth]{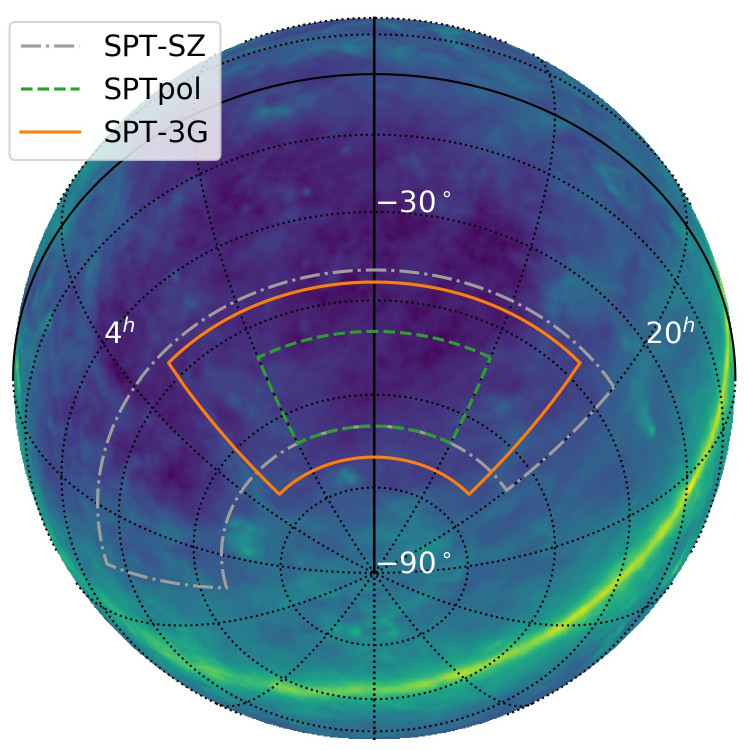}
  \end{center}
  \caption{Orthographic projection of the SPTpol 500\,$\mathrm{deg^2}$(green, dashed), SPT-SZ 2500\,$\mathrm{deg^2}$(gray, dot-dashed) and the SPT-3G 1500\,$\mathrm{deg^2}$(orange, solid) survey fields, where the background is the Planck dust map~\citep{planck13-11}.}
  \label{fig:polmap}
\end{figure*}

The 500\,$\mathrm{deg^2}$ SPTpol survey consists of $\sim$3500 single $\sim$2-hour observations over four observing seasons (May 2013 to September 2016), covering $22^h$ to $2^h$ in R.A.~and -$65^\circ$ to -$50^\circ$ in decl.~as seen in Figure~\ref{fig:polmap}.
One observation of the survey field consists of either 106 or 109 constant-elevation scans, with the telescope first scanning right and then left. After each pair of scans, the telescope takes an elevation step of $\sim$9' before making another set of paired scans. The complete observing strategy can be found outlined in~\citet{henning18}. This process is repeated until the entire 500 $\mathrm{deg^2}$ field is observed, and this set of scans serves as a single observation.

\subsubsection{Calibration Observations}\label{sec:cal_ops}
In addition to field observations, the telescope records calibration data. We observe RCW38, an HII region in the Milky Way with a known flux, once every 36-hour refrigerator cycle. We use these observations to calibrate the detector time-streams. In addition, we take brief RCW38 measurements during breaks in our observations to estimate required pointing adjustments, as discussed in Section~\ref{sec:point}. We take regular measurements of an internal calibrator, which is a heat source behind the secondary mirror, to update the calibration of our RCW38 observations on short timescales, as explained in Section~\ref{sec:relcal}. Observations taken while the telescope is fixed (noise stares) are used to characterize the noise properties of each detector.

\subsection{Map Making}
Map making refers to the process of making images of the sky from individual detector time-streams. Before this process can start some processing must be performed on the time-stream data, this includes pointing, relative calibration, and filtering. In this section, we will outline time-stream processing, pointing, calibration, and how we go from raw time-streams to maps.

\subsubsection{Time-stream Processing}\label{sec:tod}
Raw time-ordered data (TOD), also referred to as time-streams, represent the response of each detector recorded at a frequency of 190.7\,Hz as the telescope scans the sky. Each detector's data is tracked over time, and the amount of power required to maintain each detector's superconducting transition bias point during observations is recorded. We apply several filters to the individual detector TOD. First, we reduce the computational needs by downsampling our time-streams by a factor of 2. We then apply a high-pass filter to eliminate any unwanted low-frequency instrumental and atmospheric noise from the data. From the time-stream of each detector, a low-order Legendre polynomial and low-frequency sinusoids are projected out during every scan, resulting in an effective high-pass filter in the scan direction with a cutoff in angular multipole of $\ell$ = 300.

Areas within a 5' radius of point sources that are brighter than 6\,mJy at 150\,GHz (including all the sources in this study) are masked in the time-streams during high-pass filtering to prevent the occurrence of unwanted ringing artifacts. 
To prevent higher-frequency noise from interfering with the signal, we apply a low-pass filter to the time-ordered data at a multipole value of $\ell$ = 20000, which helps eliminate any aliasing caused by the map pixelization. Finally, two spectral lines associated with the SPTpol cryogenic system are masked in Fourier space, along with their harmonics (up to the third). This is done by notch filtering.

\subsubsection{Pointing}\label{sec:point}
The maps used for this analysis have been pointing-corrected by comparing the observed sky positions of known, bright astronomical sources to their expected positions. We use measured pointing errors to fit for parameters in a pointing model that corrects for time-varying deformations in the telescope structure. The pointing model with updated parameters is used to create pointing-corrected maps in a subsequent process. We utilize calibration measurements of the Galactic HII region RCW38. These measurements are conducted in between every observation. For more details on this process and the pointing model used on the SPT, we refer the reader to~\citet{chichura25}.

\subsubsection{Relative Calibration}\label{sec:relcal}
The relative calibration is based on our observations of the Galactic HII region RCW38 every 36 hours as mentioned in Section~\ref{sec:cal_ops}. The calibrations are then updated using an internal calibrator source before and after every field observation. For additional information on this process, see~\citet{schaffer11} and \citet{quan26}. 

\subsubsection{Maps}
SPT's map-making process uses a biased map maker based on MASTER~\citep{hivon02} that has been reworked to include the use of polarized data. After filtering, the time-stream data undergoes a series of quality assurance cuts, which include cuts on bolometer data from individual constant-elevation scans and from complete observations. Every pixel within a map is given a weight based on the inverse of the PSD (power spectral density) of each detector that passes through that pixel. We calculate the individual PSD for each detector between 0.8 and 3\,Hz using left-minus-right scans. The PSD is calculated using left-minus-right scans because these scans remove the CMB and other astrophysical signals; thus, only noise will remain. We use these weights for map cuts, border apodization mask creation, and observation bundling. These weights and the calibrated detector readings are binned into map pixels using a flat pixelization of the sky in the oblique Lambert equal- area azimuthal (ZEA) projection. For a more detailed description of how time-streams are turned into maps, see~\citet{crites15} and~\citet{henning18}.

\subsection{Map Cuts and Bundling}
Given that maps have a given weight inversely proportional to their variance, it is tempting to keep all maps in the analysis and let the weighting manage the down-weighting of the noisier maps. This is effective for maps exhibiting the appropriate correlation between the calculated variance and the sky signal in the map, which is applicable to the majority of maps, though not all of them. This analysis excludes maps exhibiting anomalous systematics that are not well modeled by the map variance weights. 


Individual maps from our observations have non-uniform coverage due to the elevation steps and cut time stream data, leaving us with single observation maps which may not have values in every pixel. This poses a challenge for additional map filtering in Fourier space, as the maps will exhibit discontinuities around pixels that lack values. To alleviate this problem, many maps are combined to generate a map ``bundle" using the map pixel weights, producing maps with full coverage. We decided to set the cadence for map bundles at the cryogenic refrigerator recycling timescale. Every 36 hours, the refrigerator running the detectors at 250\,mK runs out of cooling power and has to be recycled, which takes about 8 hours. This implies that following every 28 hours of observation, there is a roughly eight-hour data-collecting break. Bundles match the telescope's natural observation cadence by using this observation break as their boundary. If fewer than 10 observations remain after map cuts, the bundle will have poor coverage, and that bundle and its observations are cut. This parameter for bundle cuts results in the elimination of approximately 15\% of our total bundles, reducing the count from 528 to 455 usable bundles.

\subsection{Absolute Calibration}\label{sec:abs_cal}
As discussed in~\citet{henning18}, our absolute calibration is determined by comparing SPTpol 150\,GHz maps with the 143\,GHz Planck maps~\citep{planck15-11} over the angular multipole range  $600 < \ell <1000$. Specifically, we calculate the ratio of the SPTpol 150\,GHz auto-spectrum to the cross-spectrum of SPTpol with the Planck 143\,GHz temperature map. Once the final calibrations have been measured, we multiply each flux density value measured from each observation (see Section~\ref{sec:units}) by the corresponding calibration value for each frequency.

\subsection{Bundle Calibration and Pointing Checks}\label{sec:buncal}
Each bundle map goes through the same absolute calibration procedure described in the previous section, and a given bundle map is multiplied by the ratio of the full-dataset calibration factor to the bundle calibration factor. We also check and correct per-bundle astrometry by comparing the positions of bright sources to those in the AT20G catalog \citep{murphy10}.

\section{Source Selection}\label{sec:sources}
\begin{figure}[!htb]
  \centering
  \includegraphics[width=.95\linewidth]{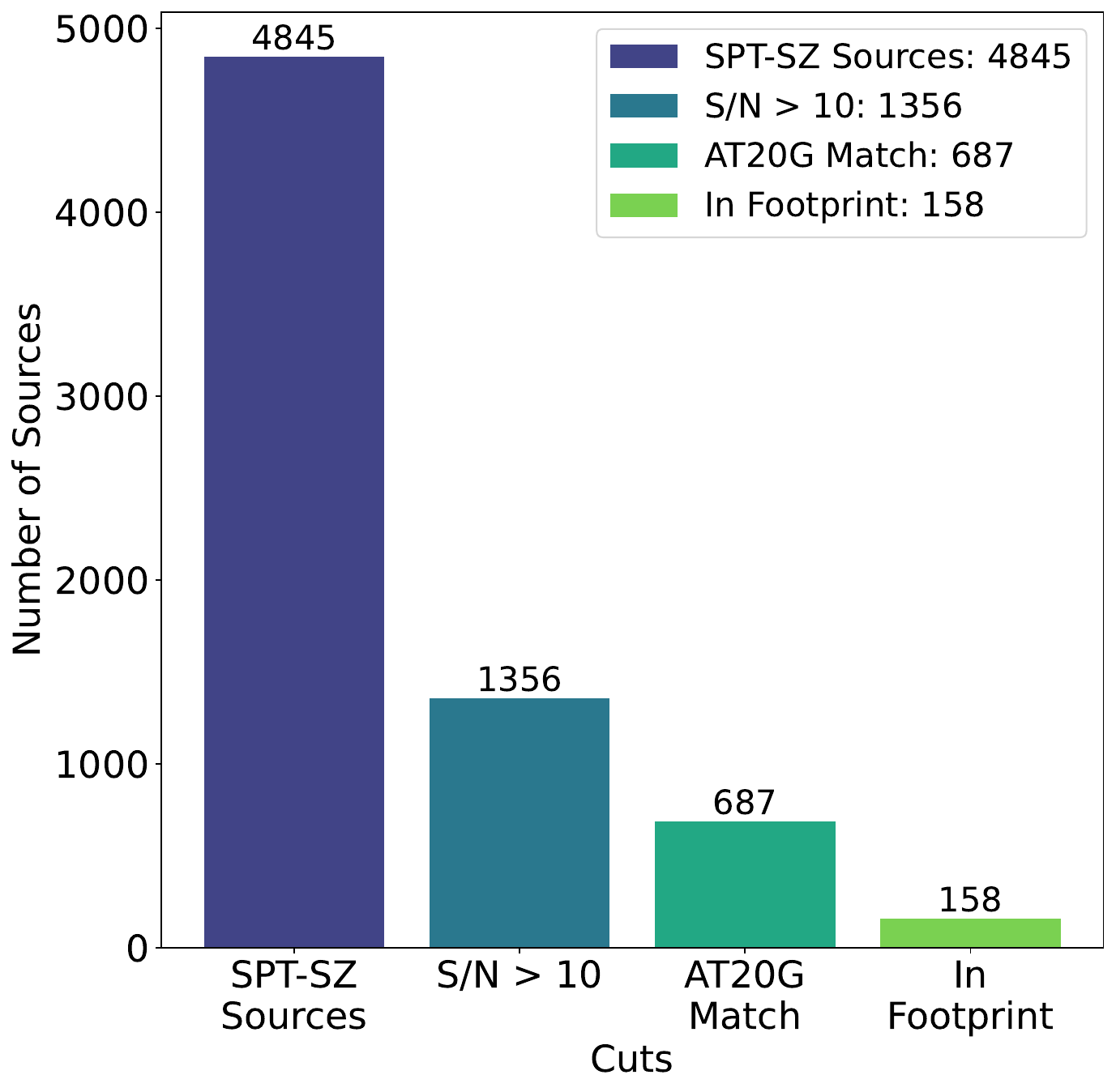}
  \caption{Source selection takes the SPT-SZ point source catalog~\citep{everett20}, extracts sources with a S/N greater than 10, cross matches the RA and DEC to the AT20G catalog, then, takes those with a pixel value with a weight value above 25\% of the maps mean weight.}
  \label{fig:selections}
\end{figure}

In this section, we detail the source selection method that leads to the establishment of our list of AGN in the SPTpol 500 $\mathrm{deg^2}$ survey field that we choose to monitor. This catalog is not meant to contain every AGN in our survey field, but rather to include all AGN that are bright enough to conduct variability studies, so the cuts have been intentionally conservative in terms of what to include. To build a list of AGN for our catalog, 
we start with the point source catalog created with the first SPT instrument, SPT-SZ. This instrument was used to conduct observations from 2007 to 2011. The primary SPT-SZ field was a 2500 $\mathrm{deg^2}$ field that spans from 20h to 7h in right ascension and from -65$^{\circ}$ to -40$^{\circ}$ in declination~\citep{story13}. As shown in Figure~\ref{fig:polmap}, the SPTpol 500 $\mathrm{deg^2}$ survey is a subset of this field. Figure~\ref{fig:selections} outlines the selection process, showing the steps we use to find and differentiate bright AGN from other point sources. The following details the step-by-step process of our AGN selection.

The initial step involves identifying all known sources in our survey field that meet our threshold for signal-to-noise (S/N) ratio. For this, we use the SPT-SZ point source catalog described in~\citet{everett20}, which contains 4845 point sources. We scale our desired S/N in a single 36-hour SPTpol 500 deg$^2$ bundle to a target S/N in the SPT-SZ catalog in the following way:
\begin{equation}
   \left(\frac{\mathrm{S}}{\mathrm{N}}\right)_\mathrm{SZ} = \frac{\sqrt{N_\mathrm{bun}}}{k \left( \frac{\mathrm{S}}{\mathrm{N}} \right)_\mathrm{bun}},
   \label{eqn:sn}
\end{equation}
where $k \approx 2$ is the ratio of noise between the map used for the SPT-SZ catalog and the SPTpol coadd, and $N_\mathrm{bun}$ ($\sim$ 450) is the number of bundled observations we have for each source. For our target of $\left(\frac{\mathrm{S}}{\mathrm{N}}\right)_\mathrm{bun} \approx 1$ at 150\,GHz, this yields $\left(\frac{\mathrm{S}}{\mathrm{N}}\right)_{\mathrm{SZ}} \approx 10$. Once we apply this cut, we are left with 1356 total sources that meet our S/N threshold. 

\begin{figure}[!htb]
  \centering
  \includegraphics[width=1\linewidth]{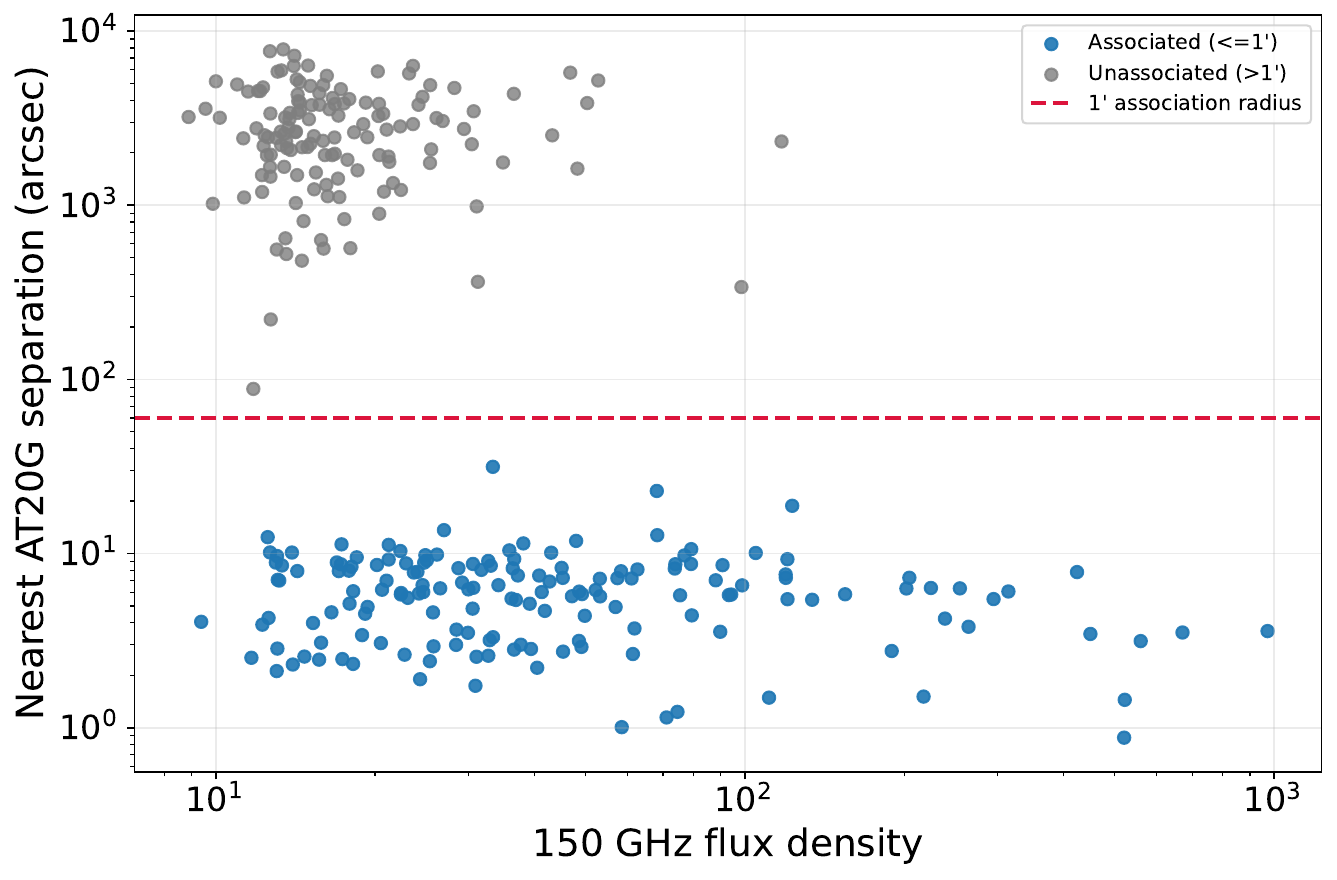}
  \caption{Angular separation to the nearest AT20G counterpart shown as a function of SPTpol 150\,GHz flux density for all SPT-SZ sources within the footprint defined by the SPTpol 25\% weight contour. The dashed horizontal line marks the 1 arcmin (60 arcsec) association radius used to define accepted matches; sources below this threshold are retained, while sources above it are treated as non-associations.}
  \label{fig:separation}
\end{figure}

We then use the AT20G catalog~\citep{murphy10} to cross-match source coordinates to within a 1' radius as seen in Figure~\ref{fig:separation}. 
%
This cross-matching aids in differentiating AGN from other point sources, such as dusty star-forming galaxies found in the SPT-SZ point source catalog---i.e., we assume every source that is above the AT20G flux density threshold of 40~mJy at 20\;GHz and above our S/N threshold in \citet{everett20} is an AGN. 

In the final step, we utilize the map weights of our 90 and 150 GHz coadds to establish a field boundary defined by the median weights of each coadd map, with the cut-off set at 25\% of the median weight value of the coadds.
To summarize, we use the following list of search parameters:
\begin{enumerate}
\setlength{\itemsep}{0pt}
  \item Does the source have a 150 GHz S/N above 10 in the SPT-SZ catalog?
  \item Does the source have a counterpart in the AT20G catalog within a 1' search radius?
  \item Is the source RA and DEC within the SPTpol 500 deg$^2$ survey footprint? Specifically, is the SPTpol map weight value at the pixel corresponding to the source RA and DEC coordinate above 25\% of the median map weight value?
\end{enumerate}

\begin{figure}[!htb]
  \centering
  \includegraphics[width=1\columnwidth]{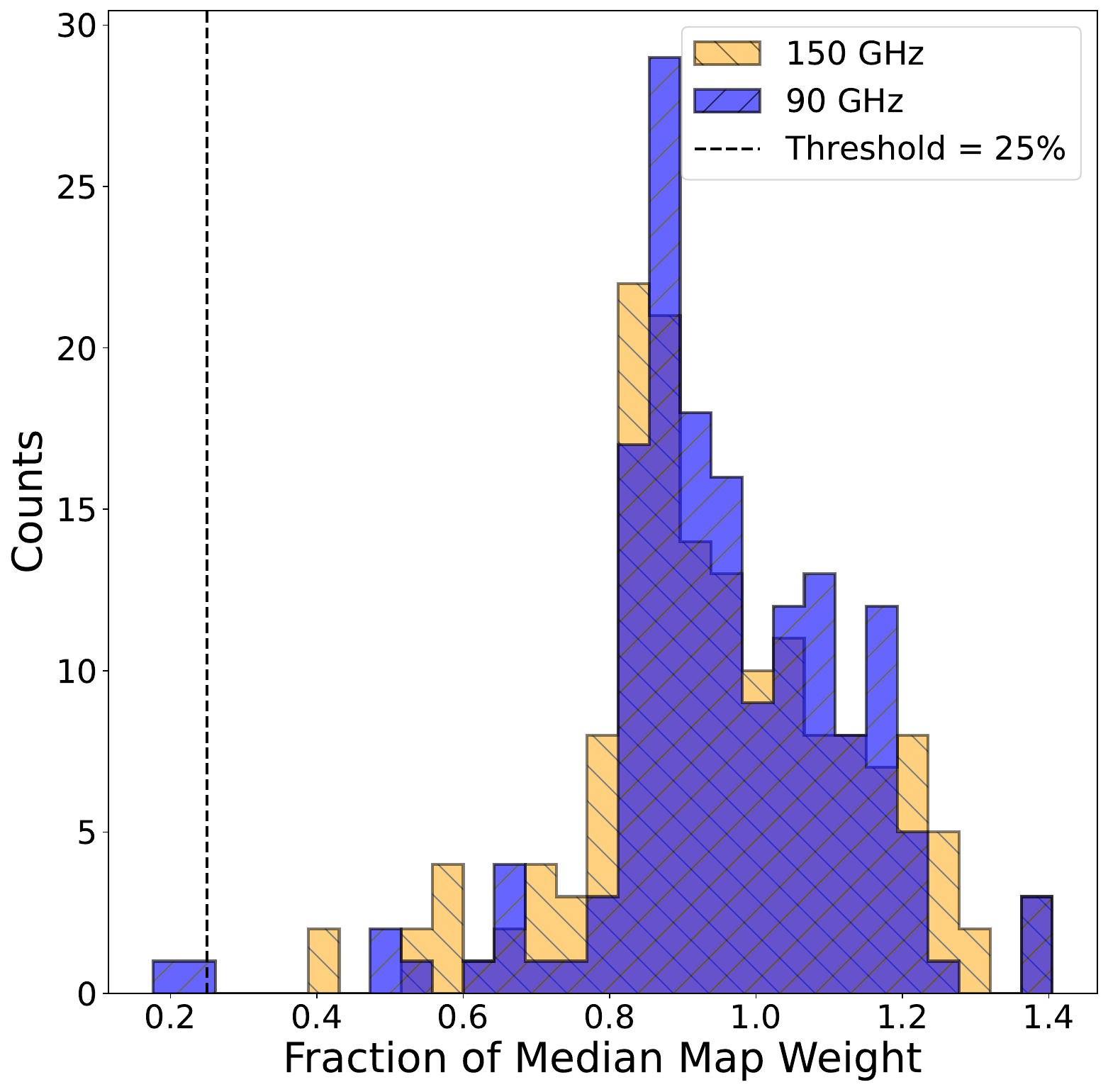}
  \caption{Fraction of median map weight values at the location of each source in the 150\,GHz coadded map (blue) and the 90\,GHz coadded map (orange). The dashed vertical line indicates the threshold for inclusion in the catalog (25\% weight).}
  \label{fig:weights}
\end{figure}

We initially applied these cuts at 150\;GHz, and we later found that applying them to our 90\;GHz data required the removal of two additional sources from the catalog. This discrepancy was due to both being on the outer edge of the 90\,GHz weight map, as seen in Figure~\ref{fig:weights}. These sources will be identified and noted on the catalog's website. Therefore we found that there were 158 AGN that fit our search parameters in the 150\,GHz dataset and 156 in the 90\,GHz dataset. The two sources that did not match all the parameters, via the 25$\%$ weight threshold in 90\,GHz, will only have their 150\,GHz band data available on the catalog website and are highlighted in Table~\ref{table:sources} to let users know which sources were affected.

\section{Building Lightcurves}
\label{sec:buildingLCs}

The next step towards making our catalog is extracting the flux density from each of the bundled maps at the location of each of our sources. In this section, we explain the steps taken to optimize the process. Section~\ref{sec:beam} describes the process of measuring the instrument beam used in our observations. Section~\ref{sec:apod} discusses the apodization mask and how it is applied to our observations. Section~\ref{sec:MFilt} discusses the matched filtering process and how we apply it to each bundle's map to bring out the signals of point sources in each bundled observation. Section~\ref{sec:units} discusses the method used to convert the observations to flux densities measured in mJy, and Section~\ref{sec:LCs} outlines how we build our light curves from the final measured flux densities extracted from our bundled observations.

\subsection{Beam Characterization}\label{sec:beam}
To perform our analysis, it is necessary to obtain a measurement of the instrumental beam, which refers to the telescope's response to a point source. We analyze the beam by taking into account seven observations of Venus, specifically by calculating the average cross-spectrum between all pairs of these observations. The Venus-derived beams are then corrected for the effect of pointing variations between individual field observations, by comparing to bright emissive sources in the full-depth field map.
Details of the beam characterization can be found in \citet{henning18} and \citet{chou25}.

\subsection{Apodization Mask}\label{sec:apod}
Although bundles have consistent coverage, their weight still falls quite sharply at the edges. Sharp features in the bundles will create problems converting between real space and Fourier space (and back again). An apodization mask serves as a tool to smooth the transitions between observed and unobserved (or masked) regions of the sky, specifically when analyzing maps in harmonic (Fourier or spherical) space. We employ a border apodization mask, sometimes known as a ``sky window function," to soften sharp map edges, similar to work discussed in~\citet{henning18} and~\citet{chou25}. We applied a threshold of 25\% of the median weight of the 90 and 150\,GHz coadd maps to help us create a different border apodization mask for each frequency. To obtain the border apodization mask, we then apply a 10 arcmin cosine taper to the edges of the binary mask.

\subsection{Matched Filtering}\label{sec:MFilt}
\begin{figure}[!htb]
  \begin{center}
   \includegraphics[width=\columnwidth]{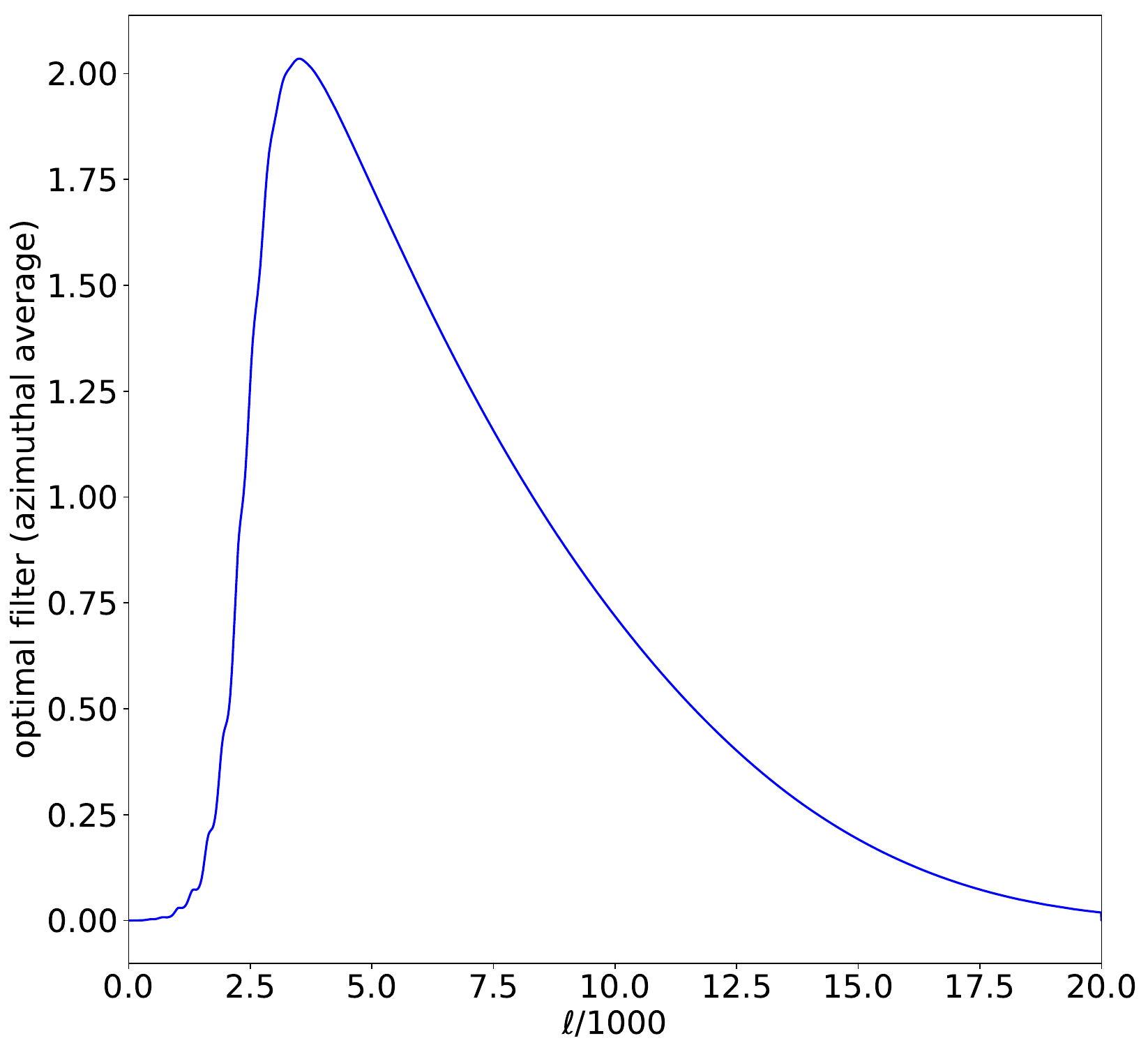}
  \end{center}
  \caption{Azimuthally averaged matched filter for point source extraction at 150\,GHz.}
  \label{fig:filter}
\end{figure}

As described in \citet{everett20}, most extragalactic objects detected by SPT appear in our maps as unresolved sources. We use a matched filter to boost the signal-to-noise of these sources. This filter suppresses angular scales where the signal-to-noise on a point source is low and boosts the total signal-to-noise on point sources in a map. The application of the matched filter in this paper is specifically designed to take into account inputs like noise and sky signal variance to extract the amplitude of a signal with a known spatial distribution as detailed in~\citet{haehnelt96}. Knowing the noise power spectrum allows us to minimize the variance of our estimator. 
Figure~\ref{fig:filter} shows the azimuthally averaged optimal/matched filter at 150\,GHz. This image shows that that our matched filter is designed to suppress large angular scale features with $\ell<$ 1000 and amplify smaller scale features in the range of $\ell$ roughly between 2,000 (wavelengths of $\sim$10') and 8,000 (wavelengths of $\sim$2'). Beyond this, at very high $\ell$ values, the noise and beam begin to dominate the signal, and the filter rolls off again.

\begin{figure}[!htbp]
  \centering
  \gridline{
    \fig{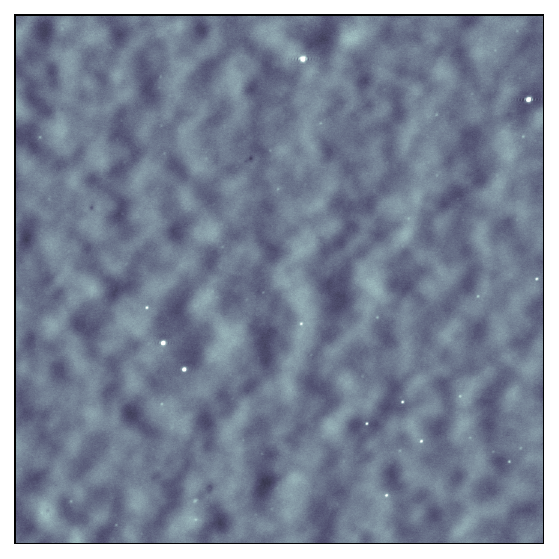}{1.0\columnwidth}{Unfiltered 5'x5' cutout of SPTpol coadd}
  }
  \vspace{-1em} 
  \gridline{
    \fig{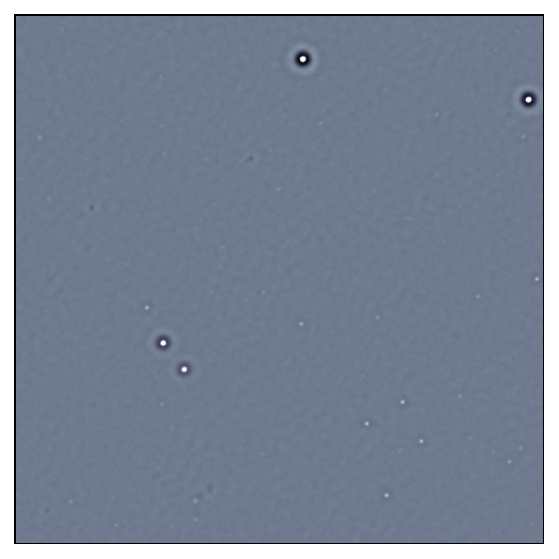}{1.0\columnwidth}{Matched filtered 5'x5' cutout of SPTpol coadd}
  }
  \caption{Top: 5'x5' cutout of SPTpol coadd map with long wavelength features still visible. Bottom: Matched filtered version of the same map cut out shown above.}
  \label{fig:maps}
\end{figure}

We first applied the matched filter and calibrations to our full four-year coadded CMB map. In these maps, we can clearly see the typical CMB patches and structures, but if we zoom in, as shown in the top image of Figure~\ref{fig:maps}, we can see that there are point sources visible but sitting on the CMB fluctuations in the unfiltered maps. Filtering largely removes the typical CMB patches and structures from the maps. In the bottom image of Figure~\ref{fig:maps}, we can see that the filtering has worked well; we see the bright point sources as bright dots with dark rings around them, which are artifacts from the filtering process.

Using our filtered maps, we extract the flux densities for each of the bright AGN in the field identified in the process outlined in Section~\ref{sec:sources}.

\subsection{Flux Density Estimation}\label{sec:units}
After making our map bundles and applying our matched filter, the extracted "flux densities" are in units of microKelvin ($\mu$K) as customary in CMB observations and are then converted to mJy as it provides a more convenient scale for the flux densities of most astronomical radio sources. That conversion is performed using the relation:
\begin{equation}\label{eqn:conv}
\frac{\mathrm{S_{\mathrm{mJy}}}}{T_{\mathrm{\mu K}}} = \frac{1\times 10^{29}}{1\times 10^{6}} \frac{\Xi \Omega_{b} e^{x} x^{4}}{(e^{x} - 1.0)^{2}},
\end{equation}
where $x$ is a constant that contains all frequency dependencies:
\begin{equation}
    x = \frac{h}{k_{b}}\frac{\mathrm{\nu}}{T_{\mathrm{cmb}}},
\end{equation}
$T_\mathrm{CMB}$ = 2.725\,K, c = $2.9979\times 10^{8}$\,m/s, h = $6.62607\times 10^{-34}$\,J/s, $k_{b}$ = $1.38065\times 10^{-23}$\,J/K, $\nu$ = $149.34\times 10^{9}$\,Hz or $96.24\times 10^{9}$\,Hz (the effective band centers discussed in Section~\ref{sec:inst}), and $\Omega_{b}$ is the solid angle of the beam measured on Venus as described in Section~\ref{sec:beam}. All of the sources that are included in this study were masked during the TOD filtering mentioned in Section~\ref{sec:tod}, so we only use the beam in our matched filter.

We also calculate $\Xi$, which is a prefactor necessary to make the conversion that includes everything but frequency dependence and the beam using the same physical constants mentioned before:
\begin{equation}
    \Xi = 2.0 k_{b} \Bigg(\frac{k_{b} T_\mathrm{cmb}}{hc}\Bigg)^{2}.
\end{equation}
%

\subsection{Light Curves}\label{sec:LCs}
Once all of the source flux densities and flux density errors are converted to mJy, the data is then saved for each individual observation and tabulated with the flux, date and errors which are then used to build our light curves. As an extra check, we made light curves for some of the brightest sources and searched for any clear correlated structure that would point to instrumental systematics as outlined in \citet{hood23}. Here, we found no evidence of similar behaviors between any of the light curves, which indicates that any fluctuations seen in the light curves are intrinsic to the sources themselves.

\section{Catalog Description}\label{sec:description}
\begin{figure}[htbp]
  \centering
  \gridline{
    \fig{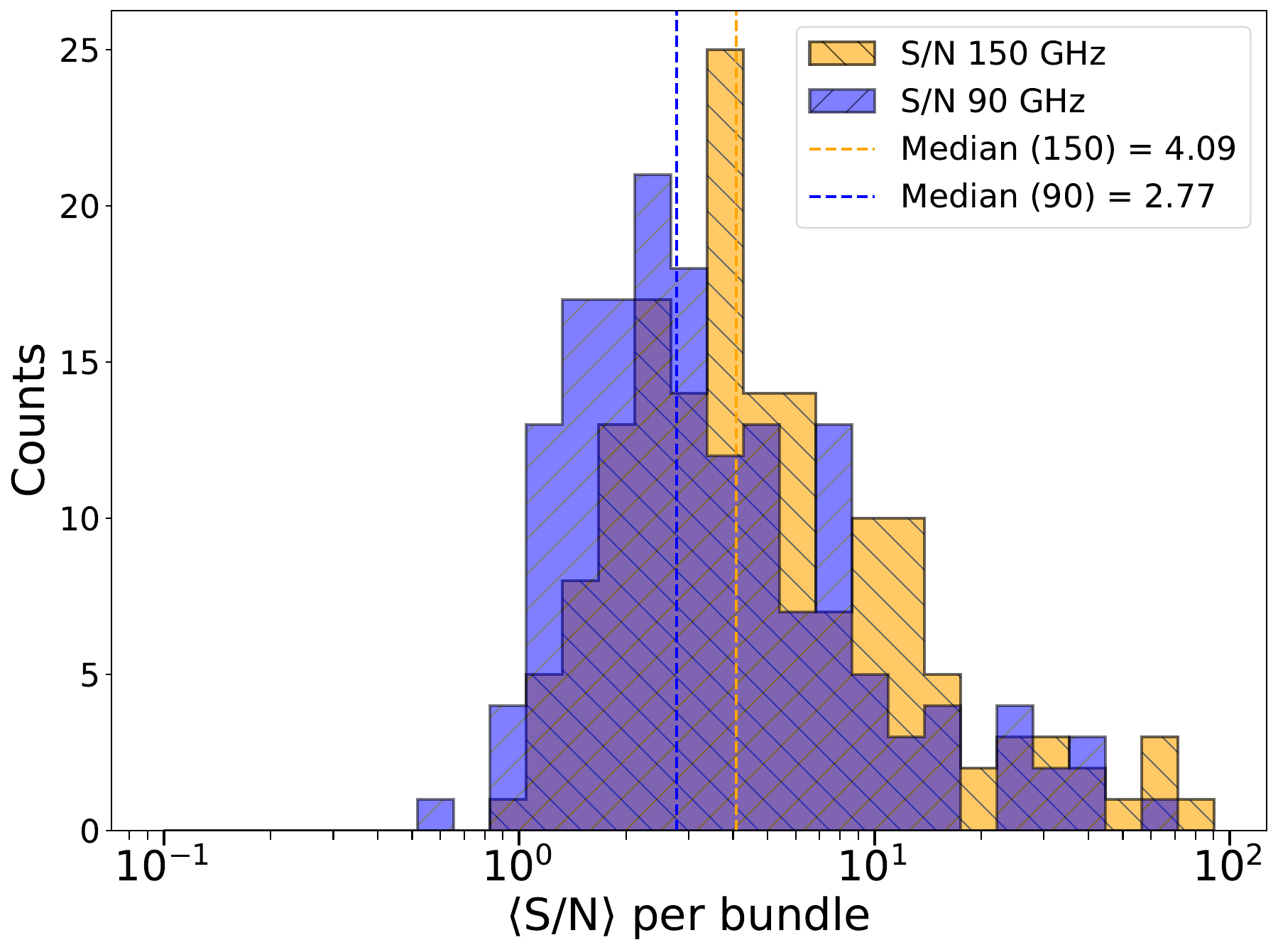}{0.9\columnwidth}{}}
  \vspace{-1em} 
  \gridline{
    \fig{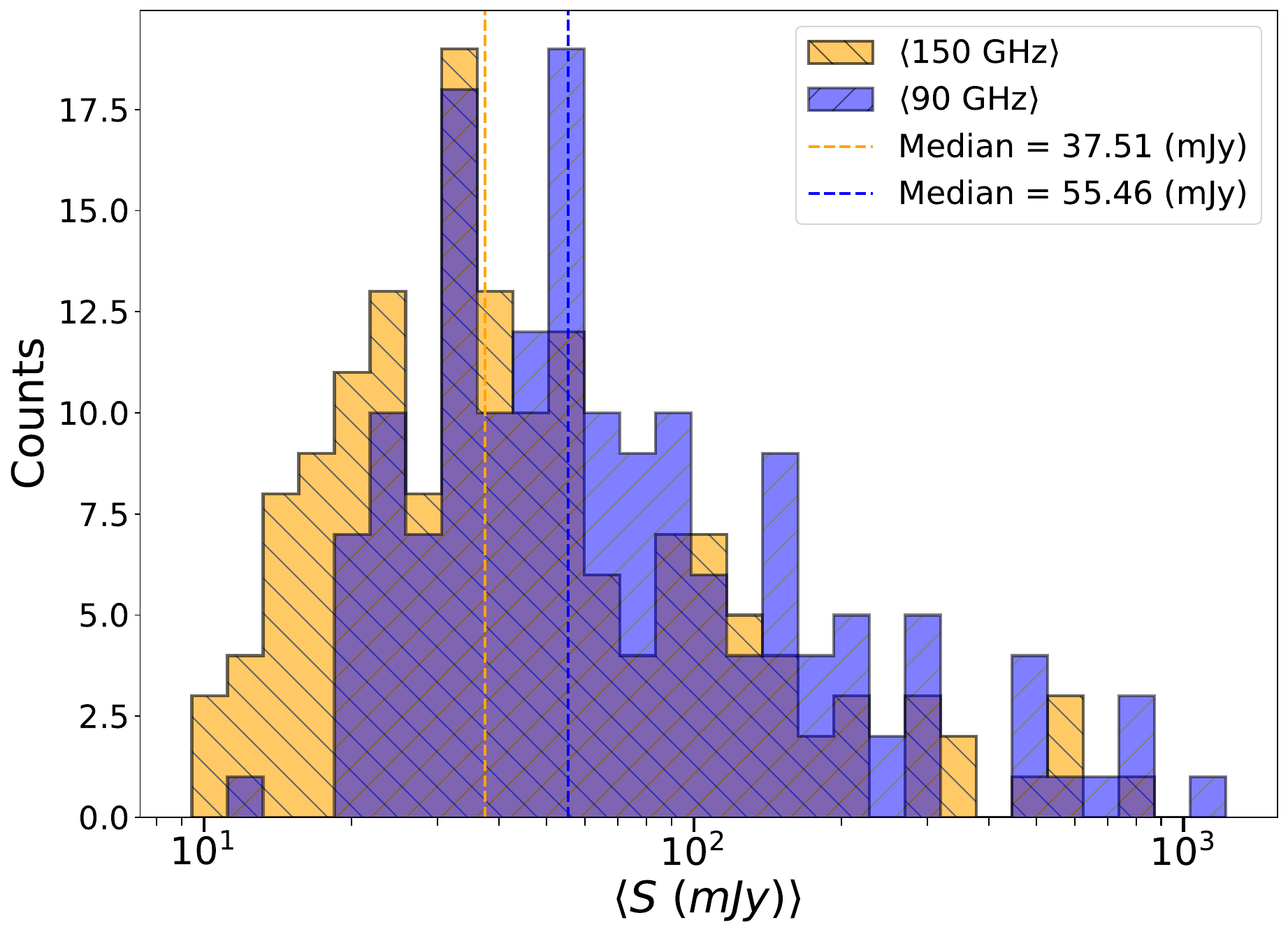}{0.9\columnwidth}{}}
  \caption{Top: Distribution of source four-year average S/N values per bundle in both 90\,GHz (blue) and 150\,GHz (orange) bands. Bottom: Distribution of mean flux densities for each source.}
  \label{fig:flux_noise}
\end{figure}

Our two-band catalog contains light curves for 158 bright AGN with a S/N ratio of 1 or more in each bundled observation at 150\,GHz, at a 36 hour cadence using the SPTpol camera between 2012-2016. Table~\ref{table:sources}, which contains the catalog specifics, provides the following seven key details about each source: 
\begin{enumerate}
\item SPT I.D.: Source IAU identification
\item RA: Right ascension (J2000) in degrees
\item DEC: Declination (J2000) in degrees
\item $\langle S_{150} \rangle$: Mean value of flux density in mJy at 150 GHz.
\item $\langle S_{90} \rangle$: Mean value of flux density in mJy at 90 GHz.
\item $z$: Redshift via \href{https://ned.ipac.caltech.edu/}{NASA/IPAC Extragalactic Database (NED)}. Redshifts were available for 57 of 158 sources.
\item $\log(\nu L_\nu)$:  Log of the product of frequency ($\nu$) and luminosity ($L_\nu$). Only calculated using 150 GHz flux, and only for sources with redshifts.
\end{enumerate}
In this table, SPT I.D., RA, and DEC are all taken from \cite{everett20}. The mean flux densities ($\langle S_{150} \rangle$ and $\langle S_{90} \rangle$) and luminosities ($\log(\nu L_\nu)$) were measured through our observations. The 150\,GHz flux density estimations have a 1$\sigma$ error of $\sim$9\,mJy per bundled observation, with a measured S/N $> 1$ as seen in Figure~\ref{fig:flux_noise}. Here, the top figure shows the mean S/N of the four-year-bundled observations for each source, and the bottom figure displays the mean flux densities across both observing bands.





%

Each reported redshift value is the result of a NED query for sources with a redshift. Appendix~\ref{sec:A2} contains a table with information about all sources in our current catalog. We note that 56 of our sources have a measured redshift, with 16 of these having a measured $\log(\nu L_\nu)$ lower than 43, suggesting that the majority of our sources are consistent with blazars. In the following section, we assume that any object lacking a measured redshift is a distant source ($z \gg 0.1$), which suggests a high probability of being a higher luminosity object, and thus likely a blazar. 

The current STRAWHAT light curve catalog is now available online at \url{https://spt3g.ncsa.illinois.edu/datasets/spt_agn_lightcurves/}. 
A machine-readable file for each source includes the following information for each bundled observation for a given source:
\begin{itemize}
   \item Date: Modified Julian Dates (MJD)
   \item Flux density: milliJansky (mJy)
   \item Flux density uncertainty: milliJansky (mJy)
\end{itemize}

The current online catalog consists of downloadable data products and light curves for each source in both 90 and 150\,GHz bands. Each light curve plot includes the timeseries data dates (MJD), mean flux density (mJy), SPT ID, RA (deg) and DEC (deg). In Appendix~\ref{sec:A1}, we provide a sample of 10 light curves of some of the brightest objects in our catalog.

\section{Discussion}\label{sec:discuss}
To try and understand some of the underlying physics behind the measured emission of our catalog population, we used the multi-frequency observations of each source to answer the following questions:
\begin{enumerate}
\setlength{\itemsep}{0pt}
  \item Is there a significant correlation between the measured fractional intrinsic variance (FIV) and mean spectral index ($\alpha$)? 
  \item Do we see a bluer-when-brighter trend among the brightest of our sources?
\end{enumerate}

A key motivation for the first question stems from the assumption that AGN with flatter spectral indices represent a fundamentally different population from those with steeper spectra. If this distinction reflects underlying physical differences—such as orientation, jet composition, or accretion properties—then it is reasonable to ask whether these populations also differ in their typical levels of variability. For example, if flat-spectrum radio quasars (FSRQs) are systematically more variable than BL Lac objects, we would expect to see this in our study.

To explore this, we determine whether there is a significant relationship by measuring the correlation between FIV and SI. We interpret any observed correlations as potential evidence of an association between the AGN type and variability amplitude. Therefore, in Section~\ref{sec:corcoef}, we evaluate the significance of these differences relative to a null hypothesis in which FIV and SI are uncorrelated, ensuring that any trends we observe are not due to random fluctuations or selection effects.

The key motivation for the second question is the observation by \citet{ulrich97} that, at sub-millimeter wavelengths, blazar spectra tend to flatten when the source becomes brighter, with BL Lacs showing stronger flatter-when-brighter trends than FSRQs in that study. We seek to determine if, without knowing the specific classification of each source (such as FSRQ or BL Lac), we see any evidence for bluer-when-brighter, or, in the millimeter case, ``flatter-when-brighter" behavior in our sample of AGN.

As discussed in Section~\ref{sec:trends}, to explore this flatter-when-brighter trend, we measure the spectral index differences between bright and faint states, allowing us to statistically assess whether AGN in our sample tend to flatten or steepen during high-flux episodes. A spectral index difference distribution centered near zero would indicate no systematic trend, while a skew toward negative values would support a ``steeper-when-brighter" scenario. Conversely, a positive skew could suggest a flatter-when-brighter behavior.

\subsection{FIV vs. Spectral Index}\label{sec:corcoef}
To answer the first question, we calculated a weighted mean spectral index $\alpha_{w}$ by first calculating a more robust mean flux density and error for each source that includes a statistical weighting for the flux densities and standard deviations. We then used these values to calculate the weighted mean spectral index and determine the errors on each spectral index value. We define our weighted mean flux densities through
\begin{equation}
    \langle F \rangle = \frac{\sum_i \left( \frac{F_i}{\sigma_i^2} \right)}{\sum_i \left( \frac{1}{\sigma_i^2} \right)},
    \label{eqn:bestflux}
\end{equation}
where $F_i$ and $\sigma_i$ are the flux density and uncertainty for each individual observation. We then calculated the weighted errors for each source using
\begin{equation}
    \sigma_{w} = \sqrt{\frac{1}{\sum_i \left( \frac{1}{\sigma_i^2} \right)}},
    \label{eqn:weightedstd}
\end{equation}
and the weighted mean spectral index using
\begin{equation}
     \alpha_{w} = \frac{\log \left( \frac{{\langle F \rangle_{150}}}{{\langle F \rangle_{90}}} \right)}{\log \left( \frac{{\nu_{c150}}}{{\nu_{c 90}}} \right)}.
     \label{eqn:si}
\end{equation}
Here, the band centers $\nu_{c 90}$ and $\nu_{c150}$ are 96.24 and 149.34\,GHz respectively (calculated as described in Section~\ref{sec:inst}). Finally, we calculate the uncertainty on the weighted mean spectral index as
\begin{equation}
    \sigma_{\alpha} = \frac{1}{\log \left( \frac{{{{\nu_{c,150}}}}}{{\nu_{c,90}}} \right)} \times \sqrt{\left( \frac{{\sigma_{w,150}}}{{\langle F \rangle_{150}}} \right)^2 + \left( \frac{{\sigma_{w,90}}}{{\langle F \rangle_{90}}} \right)^2}
    \label{eqn:si_err}
\end{equation}

We calculate the FIV and the uncertainty on that quantity in a manner similar to the method discussed in~\citet{edelson02}:
\begin{equation}
    \text{FIV} = \frac{{\text{MSD}} - {\sigma^{2}}}{\langle F \rangle^2}
    \label{eqn:fiv}
\end{equation}
\begin{equation}
    \sigma(\text{FIV}) = \sqrt{\frac{2}{N_{\text{points}}}} 
    \cdot \frac{\text{MSD}}{\langle F \rangle^2}
    \label{eqn:fiverr}
\end{equation}
where $\sigma^{2}$ is the noise squared deviation (measure of the variability in our data due to noise), \text{MSD} is the mean squared deviation (the average of the squared deviations of each data point from the mean), $\langle F \rangle^2$ is the squared mean flux density for each source, and $N_\mathrm{points}$ is the number of data points in a light curve. We could impose a physically motivated prior for FIV to be greater than 0 but we choose not to because it would potentially bias the correlation measurement.
\begin{figure}
    \centering 
    \includegraphics[width=1\columnwidth]{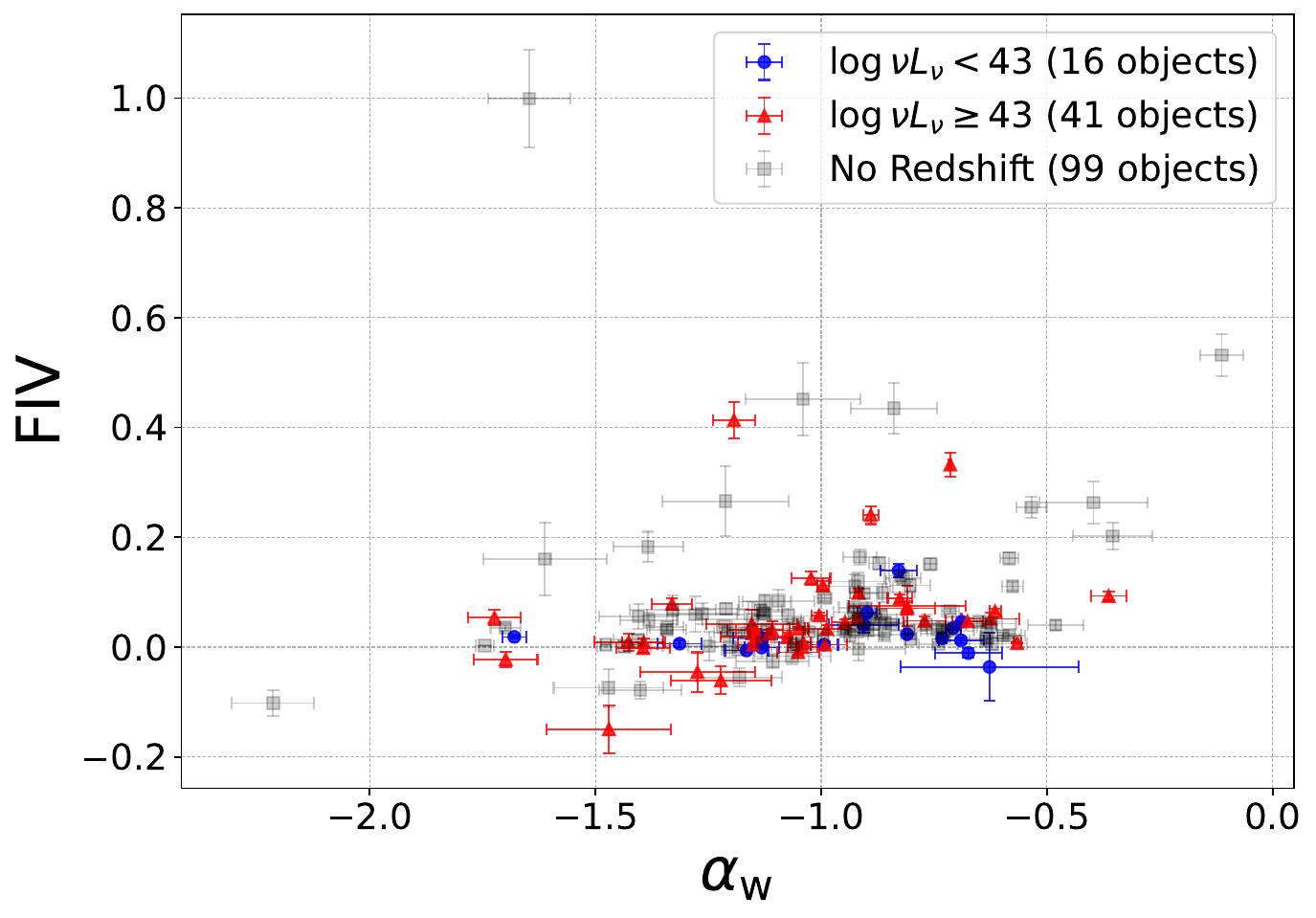}
    \caption{150\,GHz fractional intrinsic variance vs.~weighted mean spectral index. Red points indicate sources with $\log~\nu L_\nu  \geq 43$, blue points indicate sources with values lower than 43, and black points indicate sources with no measured redshift. }
   \label{fig:varvsi}
\end{figure}

Now that we have the fundamentals, we can answer the question: ``Is there a significant correlation between the measured fractional intrinsic variance and mean spectral index?" First, we measure a Spearman's rank correlation value of $\rho_\mathrm{data}=0.285$ for the 156 values of FIV and SI, shown in Figure~\ref{fig:varvsi}. To reduce the influence of known non-blazar sources, we cut the nearby/dim sources and re-compute the correlation for the subset of 139 sources with either no measured redshift or values of $\log~\nu L_\nu \geq 43$, finding a similar coefficient of 0.316. Figure~\ref{fig:varvsi} separates the sample into three groups:  Sources with $\log~\nu L_\nu  \geq 43$ are shown in red, sources with $\log~\nu L_\nu < 43$ are shown in blue, and sources with no measured redshift are shown in black.

To determine how significant this correlation is, we simulate N=10,000 realizations of our real data and measure the same data pairs for each simulation. In these simulations, the new values are centered around the real data value and allowed to vary within the measured error bars. This results in N simulated correlation coefficients, and we use the standard deviation of these values,  $\sigma(\rho_\mathrm{sims}) = 0.0261$, to determine the significance of $\rho_\mathrm{data}$. With this method, we find $\rho_\mathrm{data}$ to be 10.9$\sigma$ greater than zero. This indicates that we measure a significant positive correlation between fractional intrinsic variance and mean spectral index. 

The strong correlation found provides compelling evidence for a relationship between these variables when using spectral index as a proxy for source type. These findings motivate further investigation into the underlying astrophysical processes driving this relationship.

\subsection{Bright Source Trends}\label{sec:trends}
The ``bluer-when-brighter" trend in AGN, as discussed in, e.g., \citet{vandenberk04}, refers to the tendency of AGN, when they become brighter in optical light, to also show more blue light in their spectra. Analogously, we look for a mm-wave flatter-when-brighter trend in our sources. Such an effect is mentioned in~\citet{Robson93}, which discusses variability across the electromagnetic spectrum in AGN, specifically stating that at sub-millimeter wavelengths, blazar spectra tend to flatten when the source gets brighter.

To look for flatter-when-brighter trends in our sources, we calculate and compare the spectral indexes for the brightest half and dimmest half of each light curve of the 40 brightest sources in our sample (based on their mean four-year brightness in 90 and 150\,GHz). 
We focus on the 40 brightest sources to avoid a potential bias to the measurement of differential spectral index. Depending on how the brightest half of the light curve is defined (whether the 90 or 150\,GHz brightness, or some combination, is used), if a significant number of points in the light curve are noise-dominated, the spectral index in the bright half will reflect which band is being selected on rather than variations in the true spectral index. The 40 brightest sources all have a typical S/N of $>5$ per point, and the results of the bluer-when-brighter test on these sources is insensitive to how the brightest half of the light curve is defined.
For each of these sources, we define the quantity $\Delta\alpha$, i.e., the difference between the spectral index of the brightest half of a source's light curve and the spectral index of the dimmest half:
\begin{equation}
    \Delta\alpha = \langle \alpha_\mathrm{bright} \rangle - \langle \alpha_\mathrm{dim} \rangle
\end{equation}
\begin{figure}[!htb]
     \centering 
     \includegraphics[width=0.9\columnwidth]{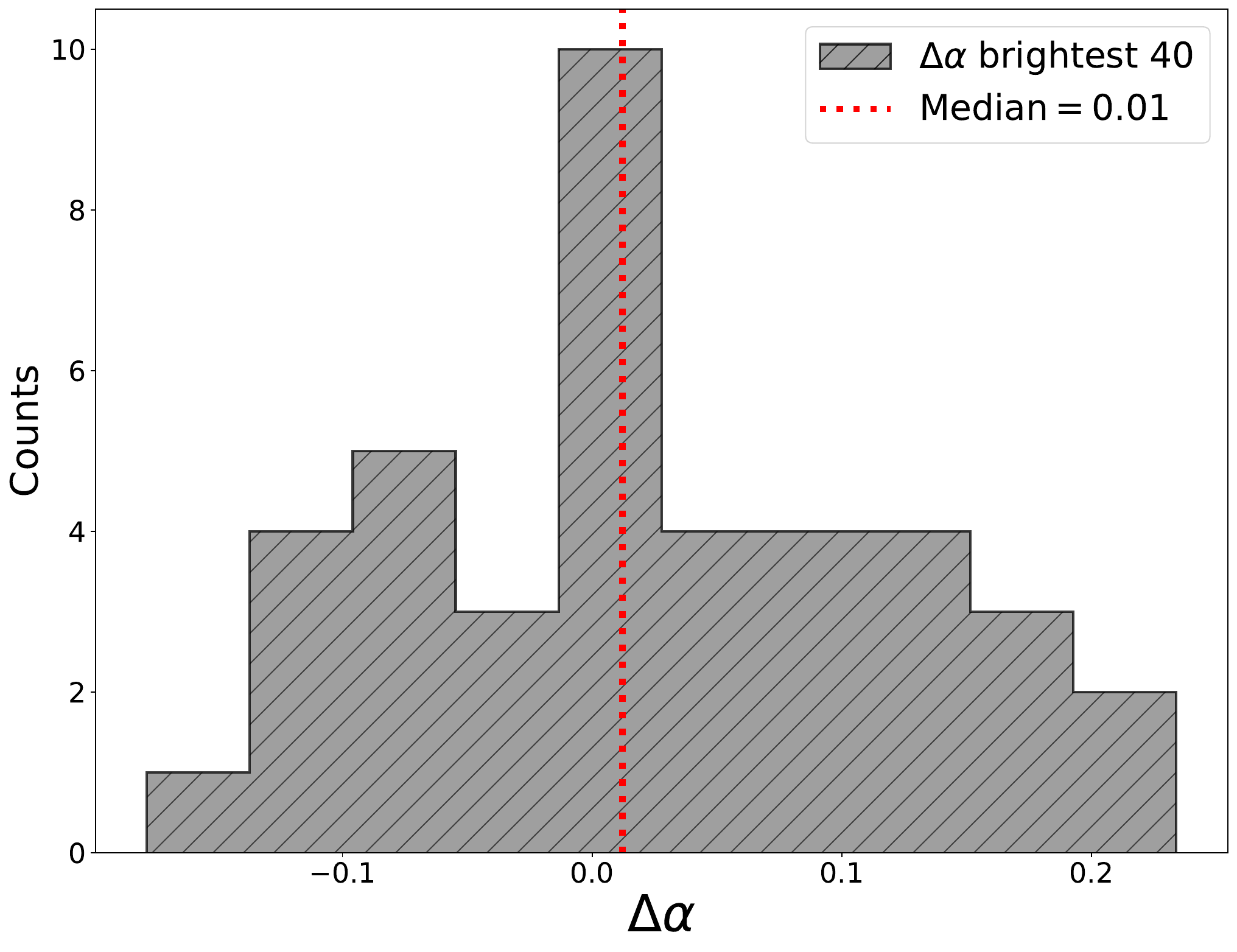}
    \caption{Delta of spectral indexes between flaring and non flaring states for each of our 40 brightest AGN light curves.}
     \label{fig:deltasi}
\end{figure}
To measure statistical significance, we calculated the mean of the $\Delta\alpha$ values and divided it by the standard error of the mean, which gave us a final significance value of 1.16$\sigma$ for the top 40 brightest sources, showing that we did not find any significant detection. 
Figure~\ref{fig:deltasi} shows that the distribution is consistent with zero, suggesting that at  millimeter wavelengths, this behavior is not a universal feature of AGN variability.


\section{Conclusion}
In this paper, we have presented a catalog of 158 AGN light curves, observed using the second-generation SPTpol instrument on the SPT. This data release marks a milestone in our AGN monitoring campaign, providing valuable insights into the variability and properties of AGN in the mm-wave band. Our analysis reveals a statistically significant (10.9$\sigma$) positive correlation between the fractional intrinsic variance and the mean spectral index, though no significant evidence for ``bluer-when-brighter" trends was found among our monitored sources. These results underscore the complexity and diversity of AGN behavior.

The provided dataset, available through the SPT Treasury Record of AGN With Historical Activity and Time-series (STRAWHAT) catalog, represents an unique and extensive resource for the astrophysical community. Researchers can leverage this dataset to further investigate AGN properties, explore correlations with multi-wavelength observations, and refine models of AGN variability. Future updates to the SPT AGN monitoring campaign will include polarization data from SPTpol and observations from the SPT-3G instrument~\citep{sobrin22}, broadening the scope of AGN studies and enhancing our comprehension of these objects. Overall, the SPT AGN monitoring campaign continues to shed light on the dynamic and multifaceted nature of AGN. We anticipate that ongoing and future observations will yield even more profound insights into the mechanisms driving AGN variability and their impact on the broader universe.

\section{Acknowledgments}
The South Pole Telescope program is supported by the National Science Foundation (NSF) through awards OPP-1852617 and OPP-2332483. Partial support is also provided by the Kavli Institute of Cosmological Physics at the University of Chicago.
Argonne National Laboratory’s work was supported by the U.S. Department of Energy, Office of High Energy Physics, under contract DE-AC02-06CH11357. 
Work at the Fermi National Accelerator Laboratory (Fermilab), a U.S. Department of Energy, Office of Science, Office of High Energy Physics HEP User Facility, is managed by Fermi Forward Discovery Group, LLC, acting under Contract No. 89243024CSC000002.
J.C.H acknowledges support from the NSF-OPP award 2219065 and the Heising-Simons Foundation.

\bibliographystyle{mnras}
\bibliography{spt_before_1995, spt_1995_to_2000,spt_2000_to_2005,spt_2005_to_2010, spt_2010_to_2015,spt_2015_to_2020,spt_2020_to_2025,spt_2025_and_after} 

\appendix

\section{Sample Light Curves}\label{sec:A1}\begin{figure}[!htb]
\centering

\gridline{
  \fig{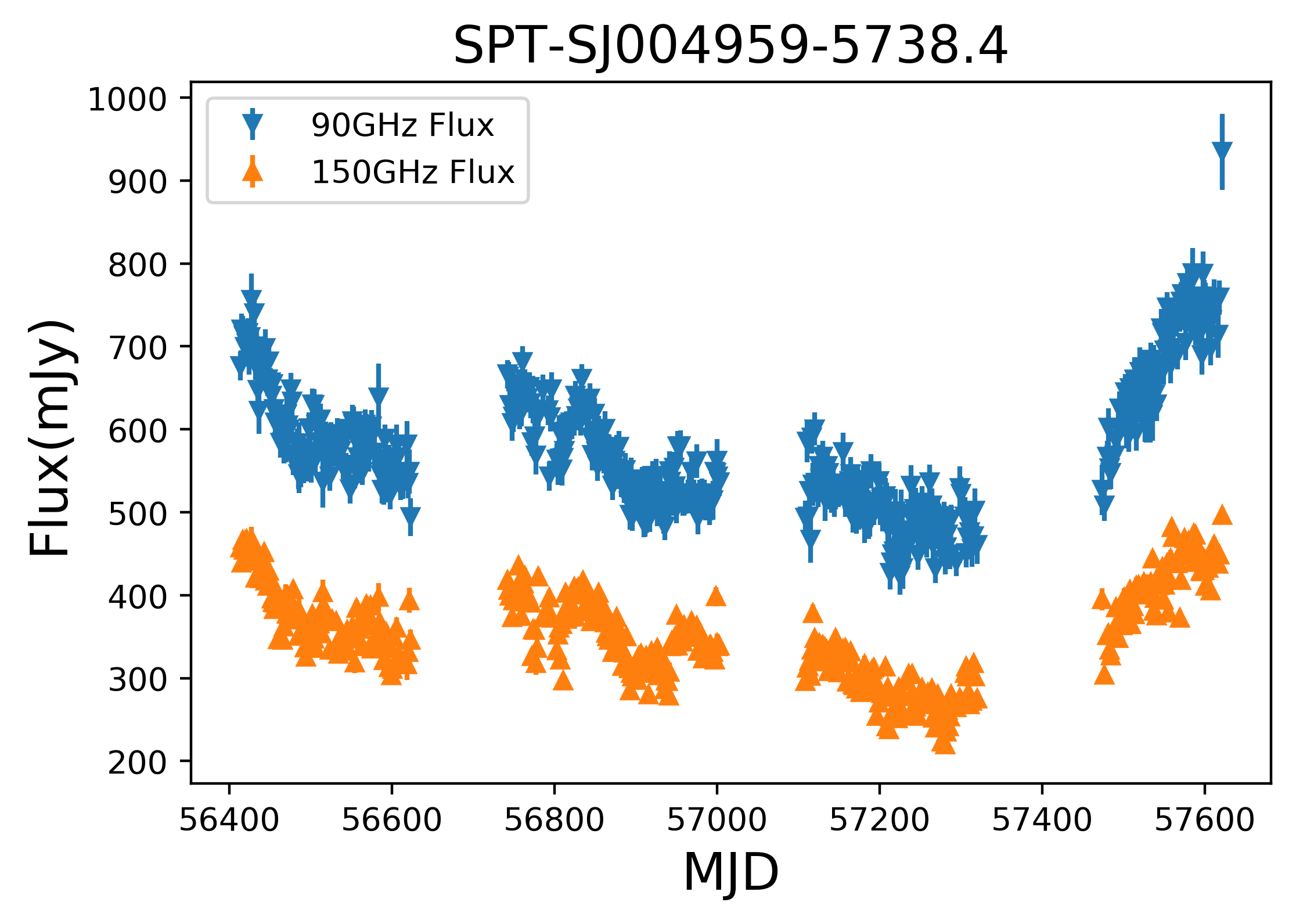}{0.45\columnwidth}{}
  \fig{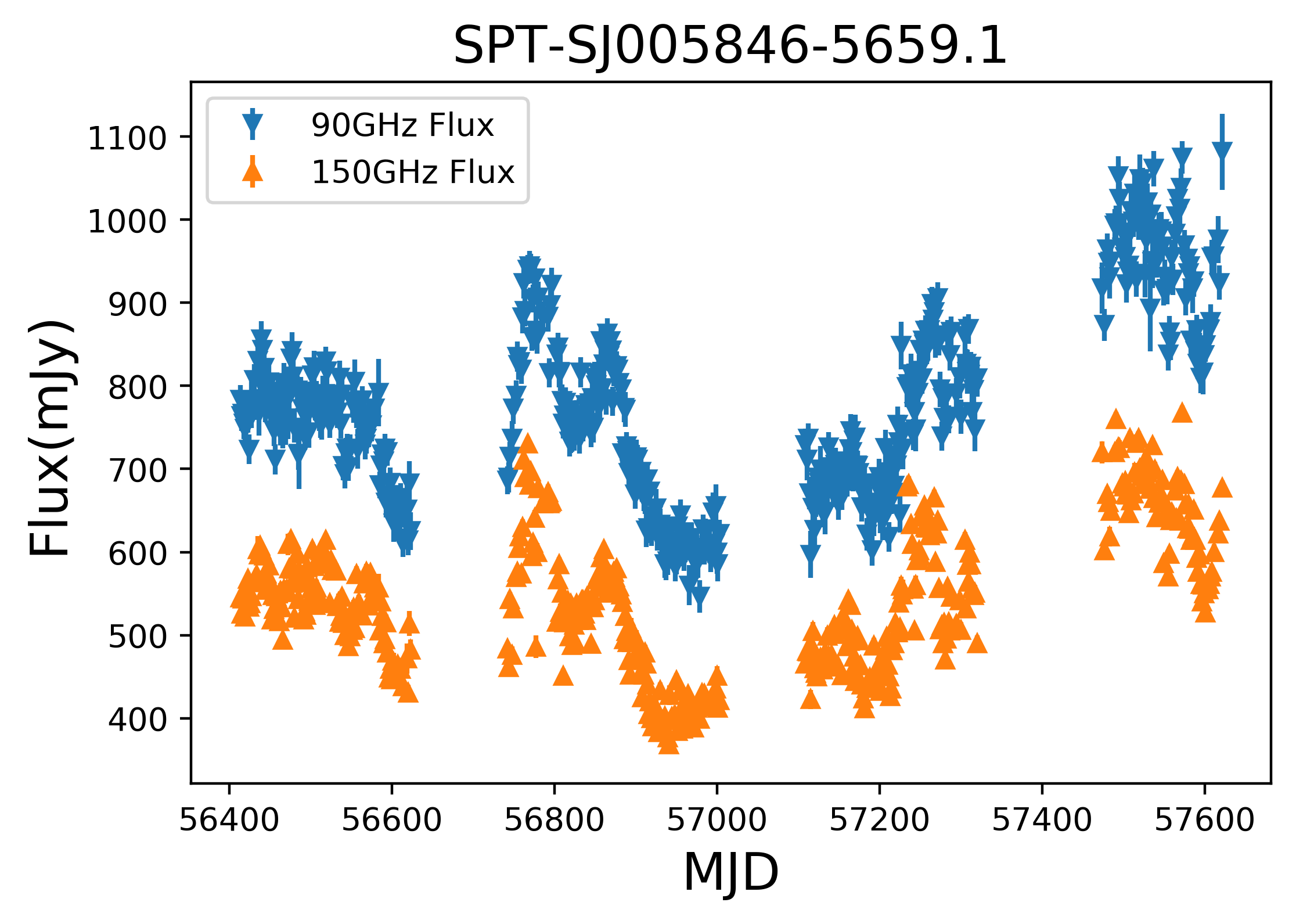}{0.45\columnwidth}{}
}

\gridline{
  \fig{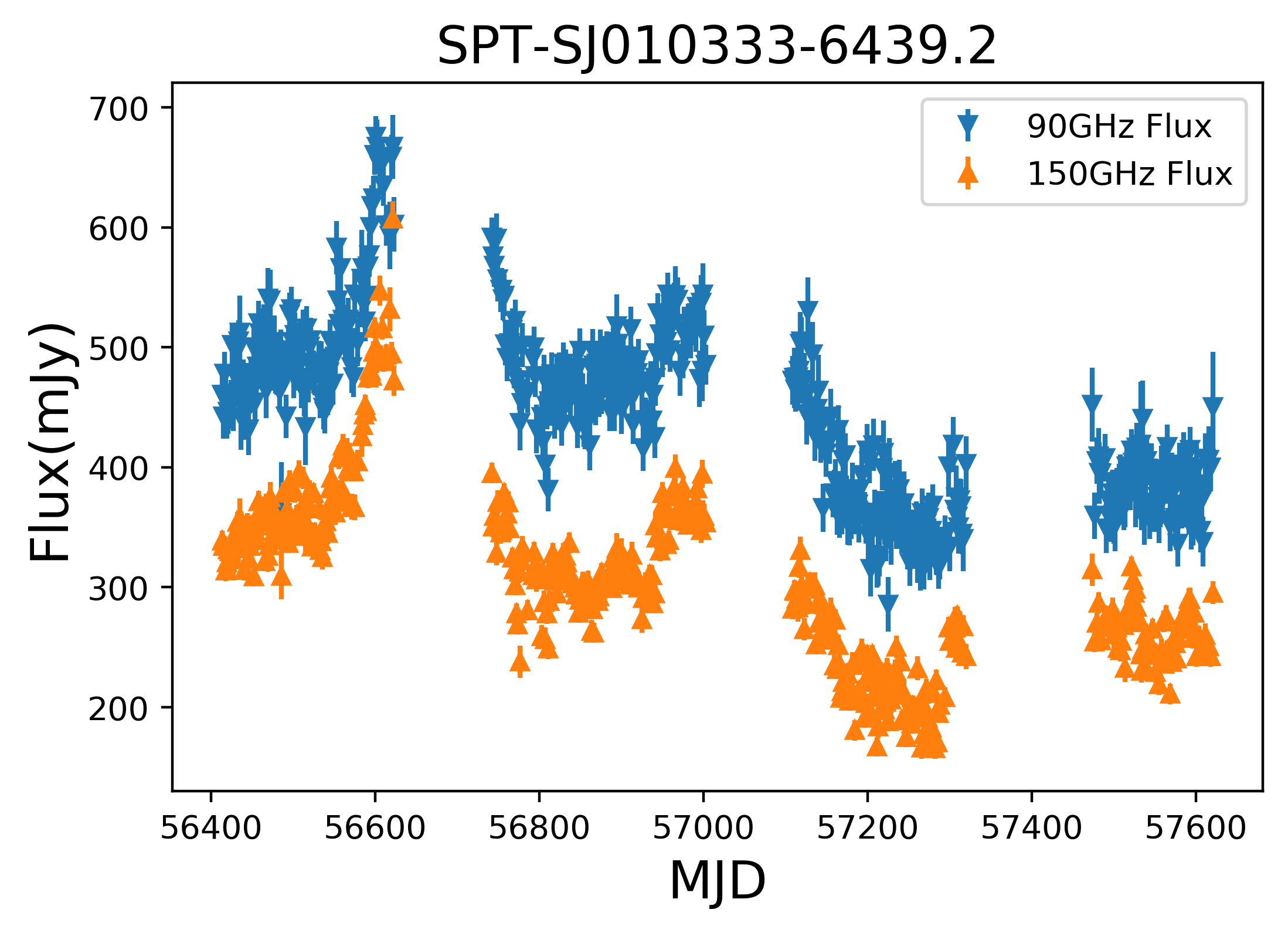}{0.45\columnwidth}{}
  \fig{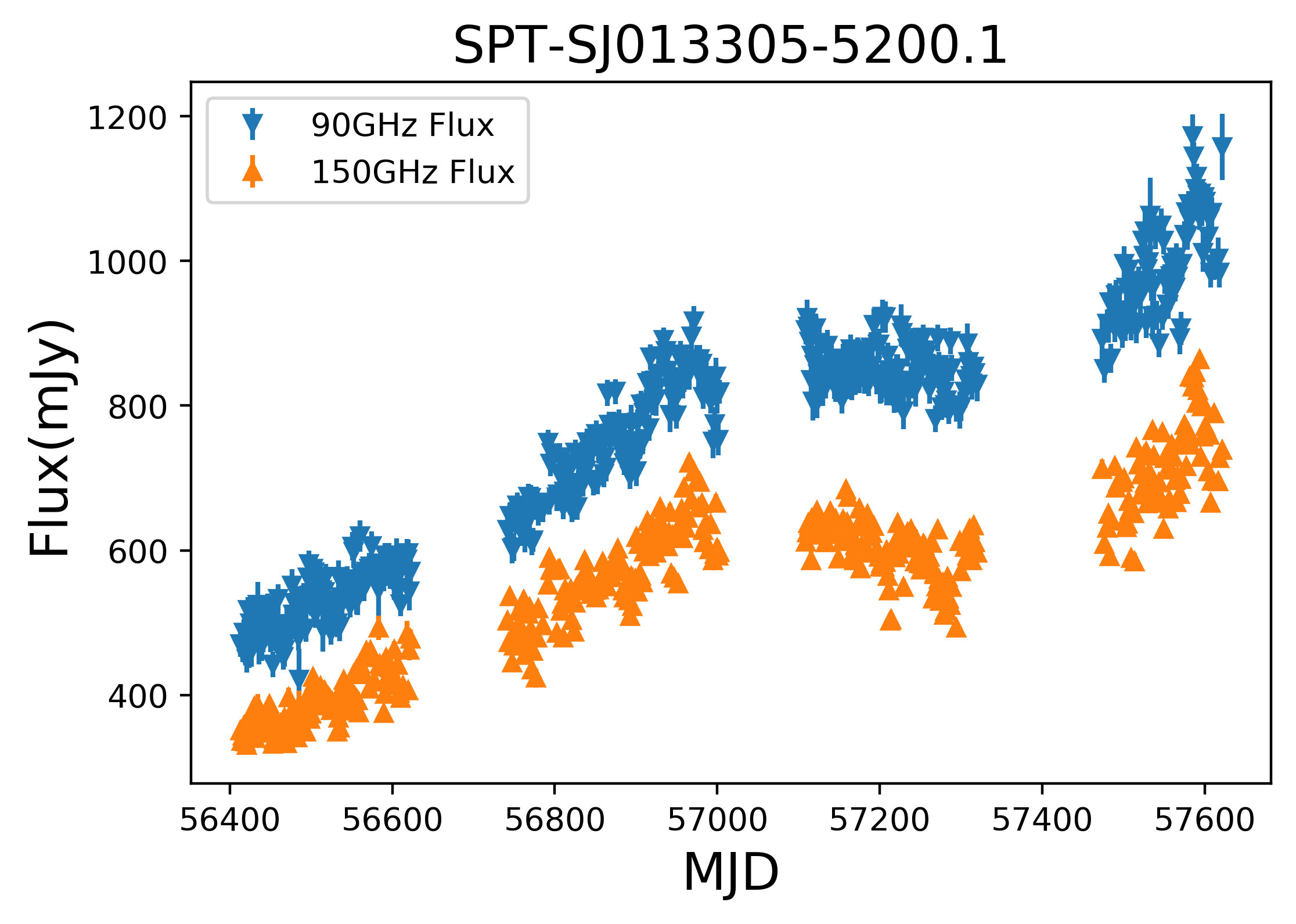}{0.45\columnwidth}{}
}

\caption{Sample 90 and 150 GHz light curves from the STRAWHAT catalog.}
\label{fig:sample_lc_1}
\end{figure}

\begin{figure}[!htb]
\centering 

\gridline{
  \fig{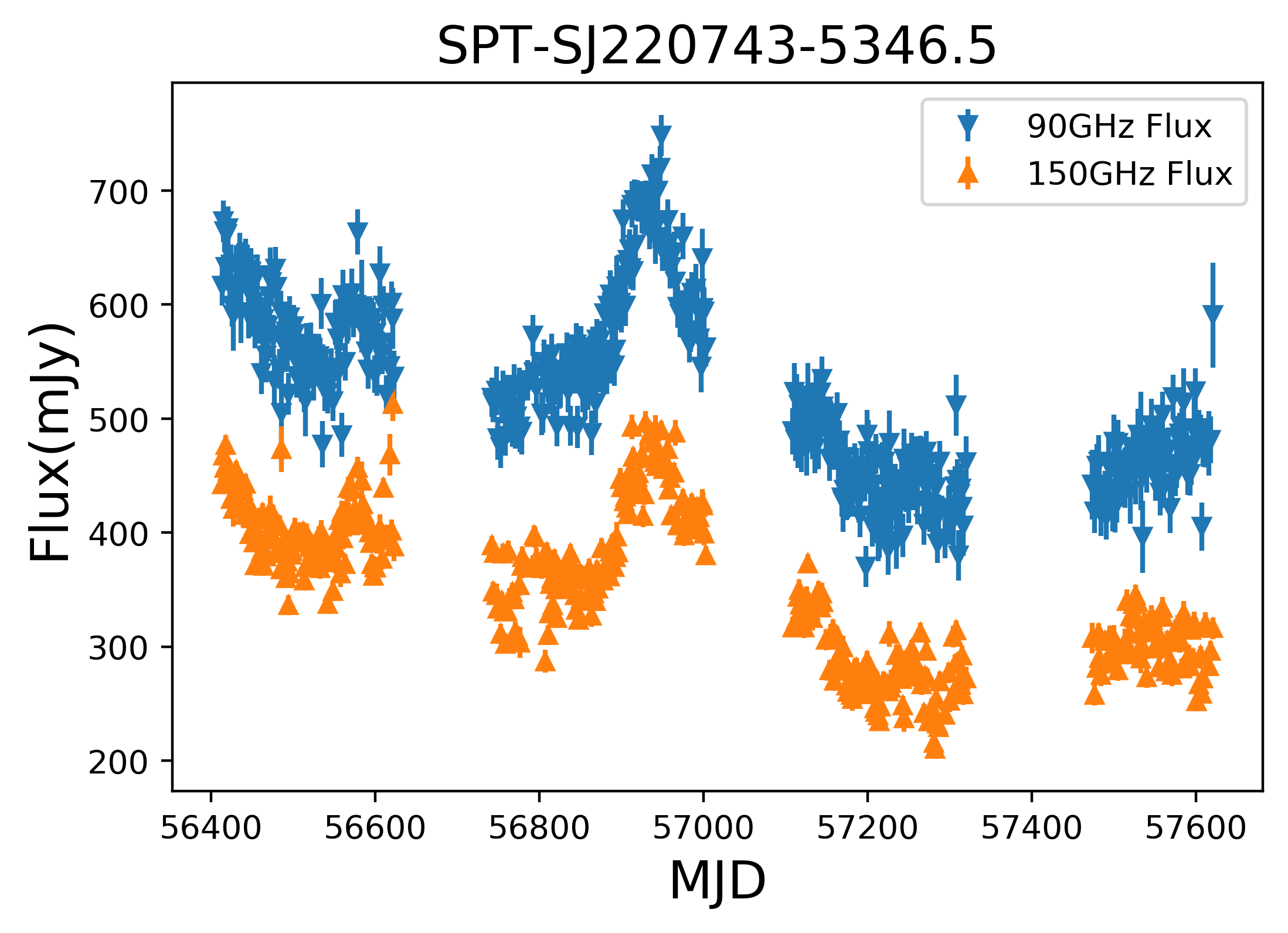}{0.45\columnwidth}{}
  \fig{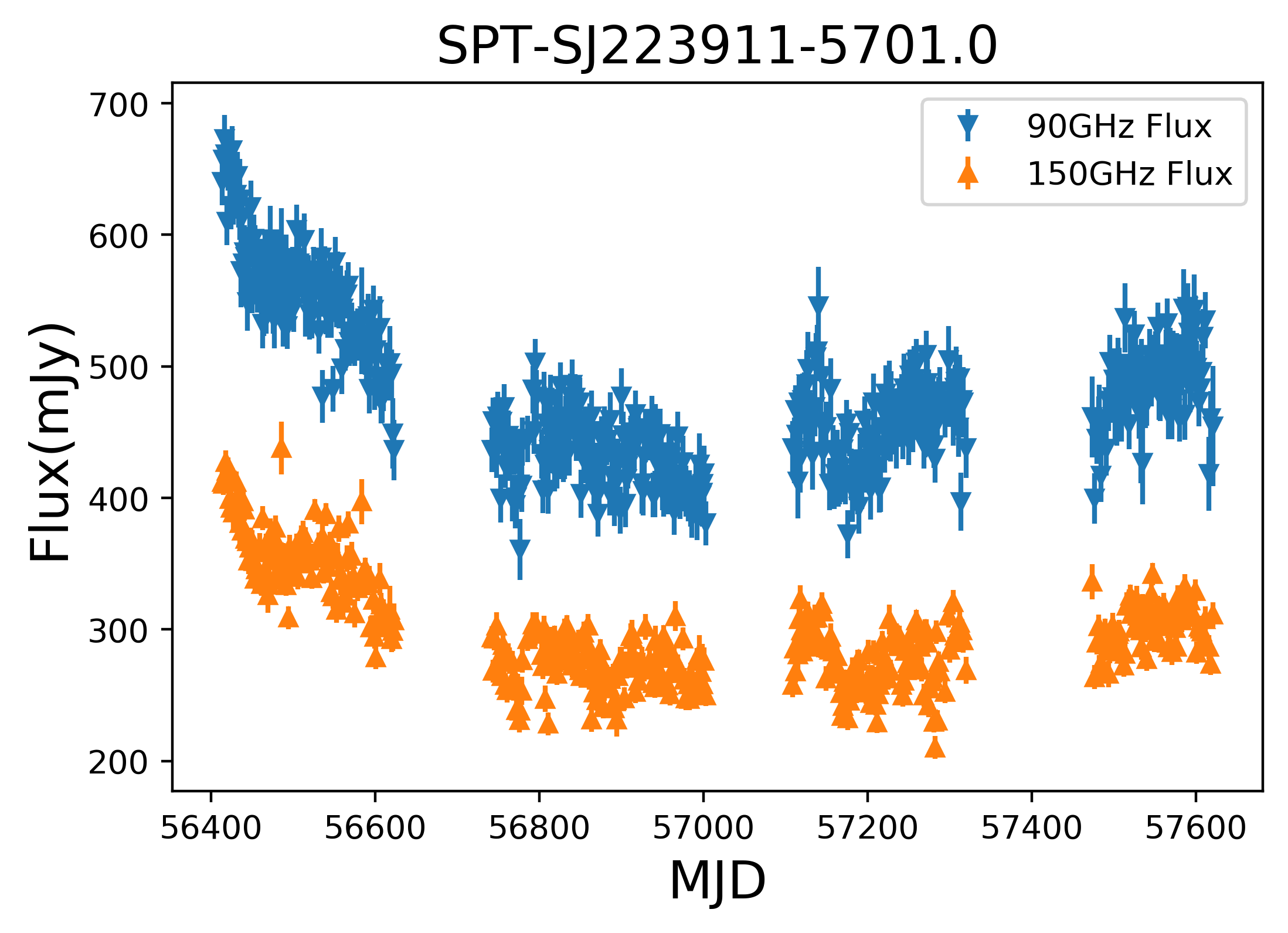}{0.45\columnwidth}{}
}

\gridline{
  \fig{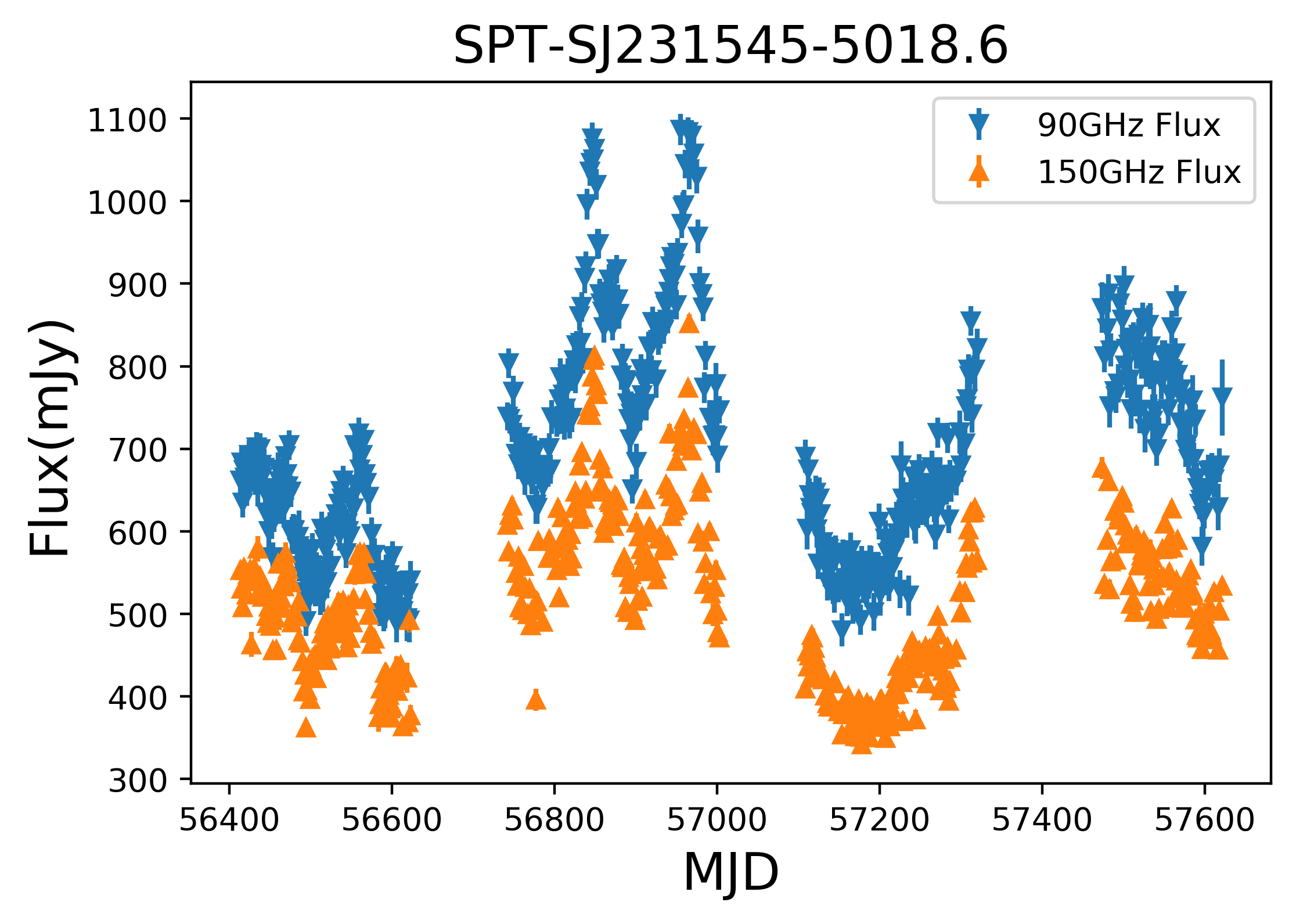}{0.45\columnwidth}{}
  \fig{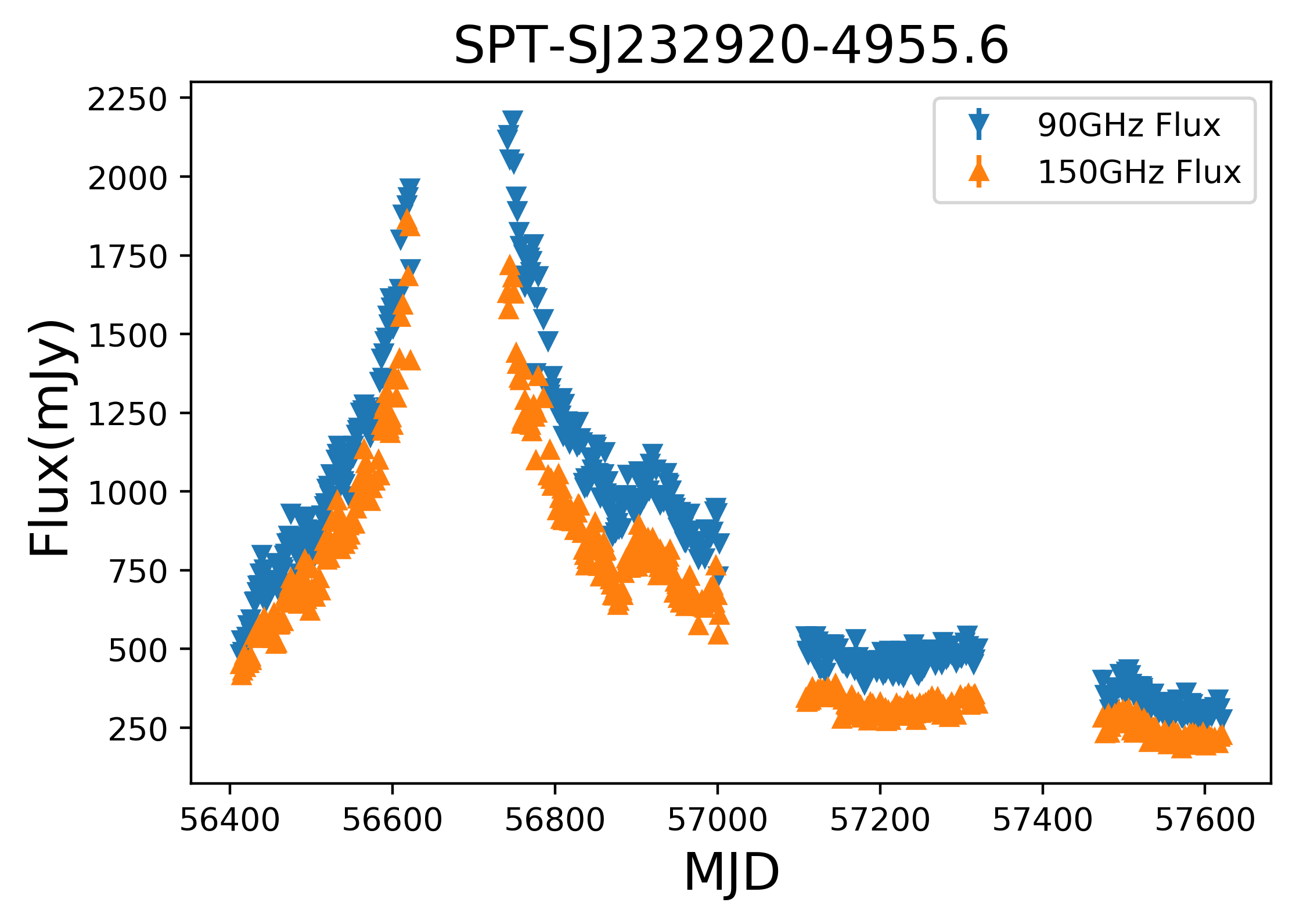}{0.45\columnwidth}{}
}

\gridline{
  \fig{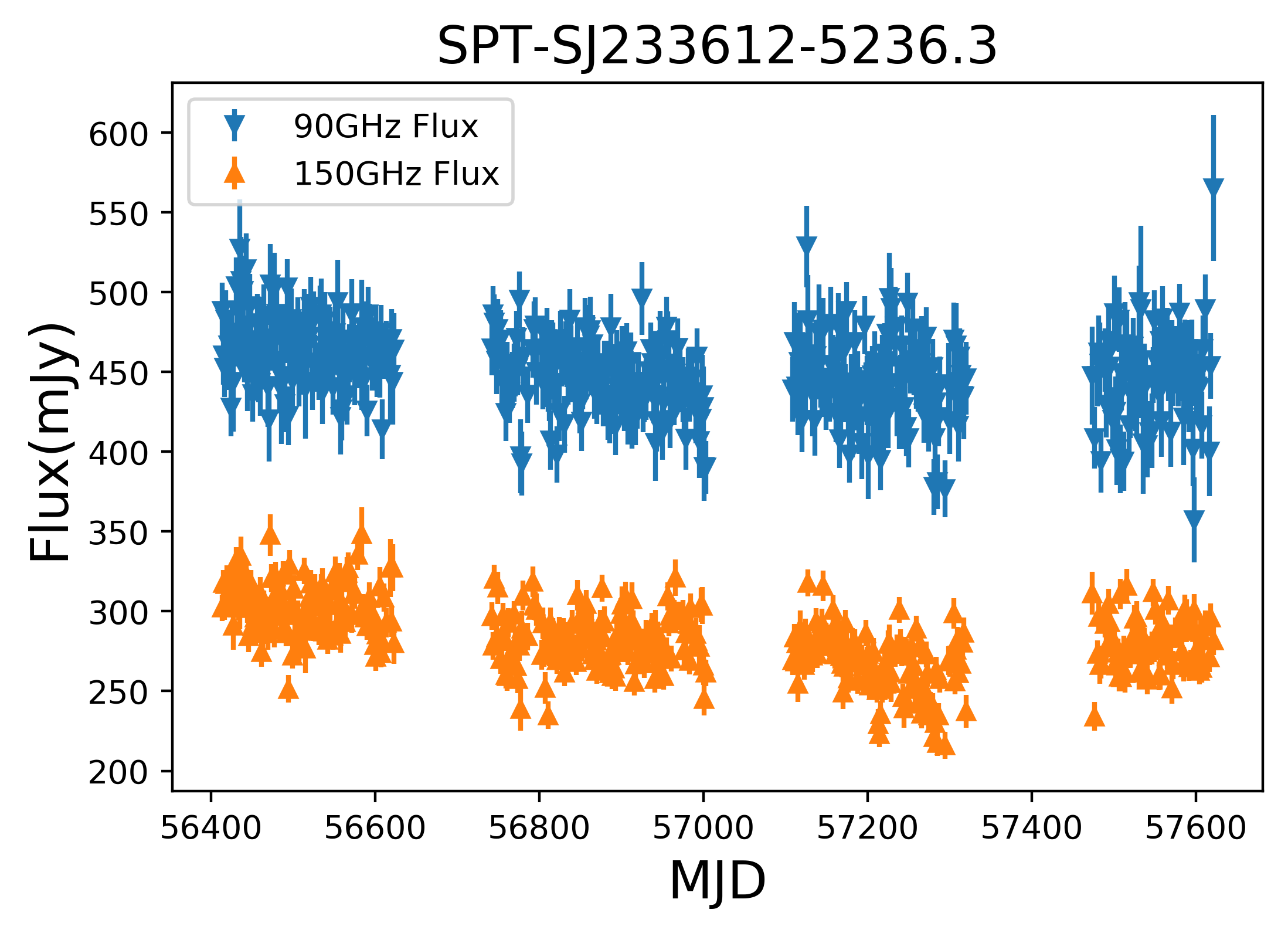}{0.45\columnwidth}{}
  \fig{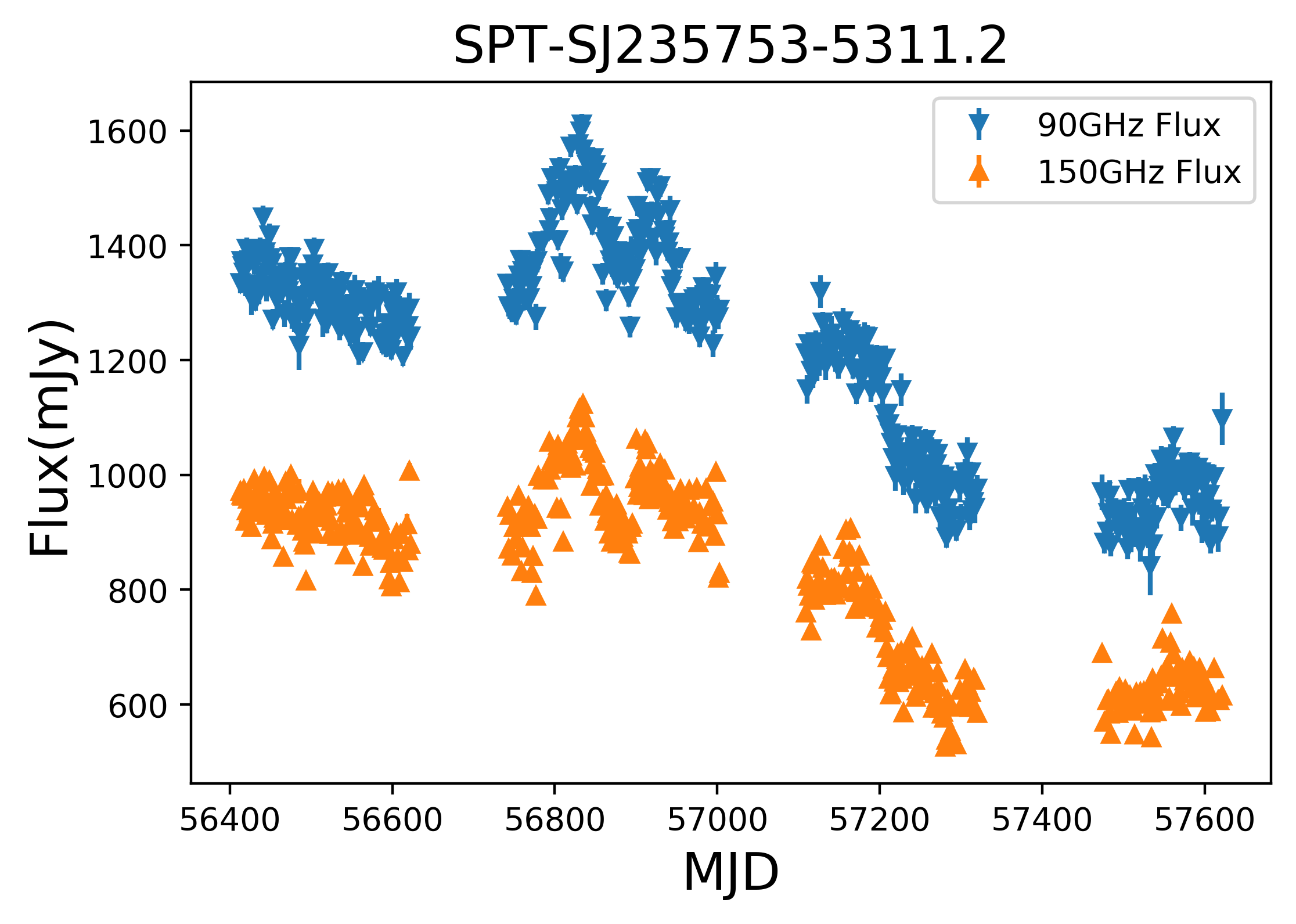}{0.45\columnwidth}{}
}

\caption{Sample 90 and 150 GHz light curves from the STRAWHAT catalog.}
\label{fig:sample_lc_2}
\end{figure}

\clearpage

\section{Source Table}\label{sec:A2}

\begin{center}
\begin{longtable}{|c|c|c|c|c|c|c|}
    \caption{List of sources with details mentioned in Section~\ref{sec:LCs}. Typical uncertainties on flux density are $\sim 1$~mJy at 150 GHz and $\sim 2$~mJy at 90 GHz. Typical uncertainties on the SPT-derived position are $< 5^{\prime \prime}$.
    The two highlighted sources are the ones excluded from our statistical analysis.}\\
    \hline
    SPT ID & RA [deg] & DEC [deg] &  $\langle S_{150} \rangle$ [mJy]  & $\langle S_{90} \rangle$ [mJy] & z & $\mathrm{log}~\nu L \nu$ \\
    \hline
    \endfirsthead
    
    \multicolumn{7}{c}%
    {{\bfseries Table Continued from Previous Page}} \\
    \hline
    SPT ID & RA [deg] & DEC [deg] &  $\langle S_{150} \rangle$ [mJy]  & $\langle S_{90} \rangle$ [mJy] & z & $\mathrm{log}~\nu L \nu$  \\
    \hline
    \endhead
    
    \hline \multicolumn{7}{|c|}{{Continued on next page}} \\ \hline
    \endfoot
    
    \hline
    \endlastfoot

\hline
        SPT-SJ000252-5948.0 & 0.71984 & -59.801495 & 79.8 & 117.0 & ~ & ~ \\ \hline
        SPT-SJ000303-5529.9 & 0.763956 & -55.499695 & 19.0 & 30.6 & ~ & ~ \\ \hline
        SPT-SJ000310-5444.7 & 0.792749 & -54.746128 & 32.8 & 54.7 & 0.0326 & 41.1 \\ \hline
        SPT-SJ000312-5905.7 & 0.803778 & -59.095818 & 16.4 & 30.3 & ~ & ~ \\ \hline
        SPT-SJ000413-5255.0 & 1.057631 & -52.917042 & 24.5 & 33.0 & 0.032 & 41.0 \\ \hline
        SPT-SJ000558-5628.5 & 1.492435 & -56.475521 & 35.0 & 55.2 & 0.291 & 43.3 \\ \hline
        SPT-SJ000719-6113.2 & 1.83149 & -61.220985 & 47.9 & 79.0 & 0.857 & 44.7 \\ \hline
        SPT-SJ000800-5243.6 & 2.002655 & -52.727661 & 81.6 & 105.0 & ~ & ~ \\ \hline
        SPT-SJ000922-5130.2 & 2.341976 & -51.504559 & 25.3 & 41.8 & 0.117 & 42.2 \\ \hline
        SPT-SJ000936-6357.5 & 2.403096 & -63.959072 & 33.1 & 52.6 & ~ & ~ \\ \hline
        SPT-SJ001324-5005.2 & 3.351625 & -50.087742 & 67.1 & 94.7 & ~ & ~ \\ \hline
        SPT-SJ001426-5638.4 & 3.612178 & -56.640419 & 34.5 & 47.5 & ~ & ~ \\ \hline
        SPT-SJ001848-5115.0 & 4.701518 & -51.250603 & 47.7 & 69.3 & ~ & ~ \\ \hline
        SPT-SJ001926-5641.7 & 4.861025 & -56.696087 & 44.0 & 63.4 & ~ & ~ \\ \hline
        SPT-SJ002233-5153.2 & 5.638746 & -51.887089 & 23.6 & 37.9 & ~ & ~ \\ \hline
        SPT-SJ002302-5650.1 & 5.759963 & -56.835518 & 53.7 & 83.1 & 0.0341 & 41.4 \\ \hline
        SPT-SJ002341-5930.5 & 5.924978 & -59.509354 & 91.2 & 118.0 & ~ & ~ \\ \hline
        SPT-SJ002510-5427.5 & 6.292719 & -54.459896 & 26.0 & 33.7 & ~ & ~ \\ \hline
        SPT-SJ002736-5409.2 & 6.903826 & -54.154541 & 9.44 & 12.1 & 0.123 & 41.8 \\ \hline
        SPT-SJ003133-5143.3 & 7.891258 & -51.722908 & 71.9 & 92.3 & ~ & ~ \\ \hline
        SPT-SJ003725-5917.9 & 9.357222 & -59.299706 & 16.6 & 27.9 & ~ & ~ \\ \hline
        SPT-SJ003735-5307.5 & 9.39682 & -53.12608 & 34.0 & 48.6 & ~ & ~ \\ \hline
        SPT-SJ003813-6112.8 & 9.555444 & -61.213886 & 12.0 & 23.0 & ~ & ~ \\ \hline
        SPT-SJ004007-5903.9 & 10.033235 & -59.066372 & 43.4 & 73.0 & ~ & ~ \\ \hline
        SPT-SJ004520-5447.7 & 11.336253 & -54.795082 & 14.7 & 25.7 & ~ & ~ \\ \hline
        SPT-SJ004905-5521.1 & 12.27424 & -55.352383 & 62.3 & 93.3 & ~ & ~ \\ \hline
        SPT-SJ004959-5738.4 & 12.497785 & -57.640041 & 354.0 & 577.0 & 1.8 & 46.5 \\ \hline
        SPT-SJ005243-4947.4 & 13.183322 & -49.791004 & 35.4 & 52.2 & ~ & ~ \\ \hline
        SPT-SJ005705-5214.3 & 14.273775 & -52.239201 & 51.1 & 67.3 & ~ & ~ \\ \hline
        SPT-SJ005846-5659.1 & 14.694412 & -56.985523 & 543.0 & 777.0 & 0.018 & 41.8 \\ \hline
        SPT-SJ005856-5219.4 & 14.734205 & -52.324444 & 70.4 & 103.0 & ~ & ~ \\ \hline
        SPT-SJ005920-6126.8 & 14.835754 & -61.447922 & 34.5 & 62.9 & ~ & ~ \\ \hline
        SPT-SJ010115-6233.2 & 15.312713 & -62.553341 & 65.3 & 105.0 & ~ & ~ \\ \hline
        SPT-SJ010210-5637.2 & 15.542074 & -56.62038 & 94.4 & 130.0 & ~ & ~ \\ \hline
        SPT-SJ010306-5109.1 & 15.775781 & -51.15247 & 53.1 & 79.4 & ~ & ~ \\ \hline
        SPT-SJ010329-5135.8 & 15.873274 & -51.597279 & 18.0 & 25.2 & ~ & ~ \\ \hline
        SPT-SJ010333-6439.2 & 15.889547 & -64.653557 & 303.0 & 446.0 & 0.163 & 43.6 \\ \hline
        SPT-SJ010350-6351.0 & 15.958658 & -63.851147 & 19.9 & 32.7 & ~ & ~ \\ \hline
        SPT-SJ010406-5657.3 & 16.02877 & -56.95536 & 43.6 & 62.3 & ~ & ~ \\ \hline
        SPT-SJ010915-6049.8 & 17.31427 & -60.830132 & 130.0 & 209.0 & ~ & ~ \\ \hline
        SPT-SJ011016-6316.0 & 17.568842 & -63.26688 & 93.6 & 158.0 & ~ & ~ \\ \hline
        SPT-SJ011323-5329.8 & 18.347488 & -53.497334 & 16.2 & 23.8 & ~ & ~ \\ \hline
        SPT-SJ011950-5357.2 & 19.961374 & -53.954586 & 122.0 & 169.0 & ~ & ~ \\ \hline
        SPT-SJ012007-5211.1 & 20.031385 & -52.185776 & 19.6 & 28.0 & ~ & ~ \\ \hline
        SPT-SJ012140-6309.2 & 20.418766 & -63.153561 & 23.3 & 48.9 & 0.834 & 44.4 \\ \hline
        SPT-SJ012456-5113.1 & 21.236553 & -51.219868 & 150.0 & 195.0 & 1.1 & 45.5 \\ \hline
        SPT-SJ012623-5103.1 & 21.598175 & -51.052017 & 24.8 & 36.1 & ~ & ~ \\ \hline
        SPT-SJ012755-5136.6 & 21.980688 & -51.610878 & 30.3 & 41.8 & ~ & ~ \\ \hline
        SPT-SJ012820-5649.6 & 22.085531 & -56.827629 & 77.2 & 114.0 & 0.0666 & 42.2 \\ \hline
        SPT-SJ012833-5255.3 & 22.139702 & -52.921925 & 16.7 & 28.3 & ~ & ~ \\ \hline
        SPT-SJ013305-5200.1 & 23.271059 & -52.001778 & 554.0 & 759.0 & 0.02 & 41.9 \\ \hline
        SPT-SJ013408-5526.3 & 23.537497 & -55.439629 & 23.1 & 34.5 & ~ & ~ \\ \hline
        SPT-SJ013540-5149.7 & 23.918295 & -51.829113 & 23.6 & 31.0 & ~ & ~ \\ \hline
        SPT-SJ013547-5244.2 & 23.947233 & -52.737869 & 19.9 & 32.3 & ~ & ~ \\ \hline
        SPT-SJ013949-5217.7 & 24.954716 & -52.296345 & 57.3 & 73.8 & ~ & ~ \\ \hline
        SPT-SJ014142-5650.6 & 25.427429 & -56.844177 & 32.7 & 54.6 & ~ & ~ \\ \hline
        SPT-SJ014225-5729.8 & 25.608095 & -57.49823 & 95.5 & 143.0 & ~ & ~ \\ \hline
        SPT-SJ014257-6044.9 & 25.737642 & -60.749836 & 42.1 & 69.6 & 0.181 & 42.9 \\ \hline
        SPT-SJ014415-6421.8 & 26.065655 & -64.364769 & 126.0 & 204.0 & ~ & ~ \\ \hline
        SPT-SJ014648-5202.5 & 26.700077 & -52.042297 & 110.0 & 149.0 & 0.0981 & 42.7 \\ \hline
        SPT-SJ015050-6044.2 & 27.709646 & -60.736839 & 23.8 & 35.3 & ~ & ~ \\ \hline
        SPT-SJ015357-5406.8 & 28.49115 & -54.114372 & 33.1 & 42.7 & ~ & ~ \\ \hline
        SPT-SJ015419-5107.9 & 28.581919 & -51.13258 & 83.5 & 132.0 & 1.58 & 45.7 \\ \hline
        SPT-SJ015509-6425.5 & 28.791323 & -64.425034 & 53.8 & 79.4 & ~ & ~ \\ \hline
        SPT-SJ015650-5439.7 & 29.211018 & -54.662025 & 41.5 & 86.9 & ~ & ~ \\ \hline
        SPT-SJ015721-5422.6 & 29.339838 & -54.377361 & 11.2 & 19.0 & ~ & ~ \\ \hline
        SPT-SJ015816-5004.3 & 29.56769 & -50.073231 & 26.6 & 39.6 & ~ & ~ \\ \hline
        SPT-SJ015826-5810.9 & 29.611347 & -58.182339 & 106.0 & 166.0 & ~ & ~ \\ \hline
        SPT-SJ015836-6411.4 & 29.650028 & -64.191368 & 50.4 & 86.3 & ~ & ~ \\ \hline
        SPT-SJ015838-6419.3 & 29.659222 & -64.322357 & 30.6 & 52.2 & ~ & ~ \\ \hline
        SPT-SJ015853-6334.8 & 29.724417 & -63.580101 & 23.7 & 43.4 & ~ & ~ \\ \hline
        SPT-SJ015912-5308.7 & 29.803917 & -53.146492 & 22.7 & 25.3 & ~ & ~ \\ \hline
        \rowcolor{yellow!30} SPT-SJ215445-5725.6 & 328.69125 & -57.427685 & 35.7 & 28.4 & ~ & ~ \\ \hline
        \rowcolor{yellow!30} SPT-SJ215543-5400.2 & 328.9295 & -54.003368 & 39.2 & 49.2 & ~ & ~\\ \hline
        SPT-SJ215648-6331.0 & 329.20093 & -63.517941 & 15.7 & 31.0 & ~ & ~ \\ \hline
        SPT-SJ220005-5455.7 & 330.02267 & -54.928963 & 120.0 & 166.0 & ~ & ~ \\ \hline
        SPT-SJ220055-5520.1 & 330.23062 & -55.335625 & 186.0 & 257.0 & 0.0489 & 42.2 \\ \hline
        SPT-SJ220145-6457.6 & 330.44019 & -64.960358 & 10.2 & 21.0 & ~ & ~ \\ \hline
        SPT-SJ220253-5635.7 & 330.72101 & -56.59528 & 37.2 & 49.3 & ~ & ~ \\ \hline
        SPT-SJ220301-4937.0 & 330.75592 & -49.617306 & 58.3 & 87.8 & ~ & ~ \\ \hline
        SPT-SJ220359-6130.3 & 330.99985 & -61.506657 & 101.0 & 158.0 & 1.21 & 45.5 \\ \hline
        SPT-SJ220715-6325.6 & 331.81311 & -63.428028 & 37.7 & 51.9 & 0.618 & 44.2 \\ \hline
        SPT-SJ220743-5346.5 & 331.93094 & -53.776051 & 351.0 & 526.0 & ~ & ~ \\ \hline
        SPT-SJ220846-6325.8 & 332.19308 & -63.430412 & 94.8 & 145.0 & 0.229 & 43.5 \\ \hline
        SPT-SJ221015-5031.1 & 332.56619 & -50.518791 & 27.9 & 40.0 & 2.77 & 46.0 \\ \hline
        SPT-SJ221032-5508.6 & 332.63431 & -55.143677 & 28.6 & 37.4 & 0.77 & 44.3 \\ \hline
        SPT-SJ221334-6329.9 & 333.39194 & -63.499844 & 40.3 & 73.0 & ~ & ~ \\ \hline
        SPT-SJ221623-5238.0 & 334.09979 & -52.633347 & 15.5 & 25.8 & 1.36 & 44.8 \\ \hline
        SPT-SJ221642-5637.5 & 334.17682 & -56.625919 & 31.9 & 39.3 & ~ & ~ \\ \hline
        SPT-SJ221818-5038.6 & 334.57654 & -50.644485 & 17.2 & 31.7 & ~ & ~ \\ \hline
        SPT-SJ221824-6454.6 & 334.60376 & -64.911026 & 39.6 & 55.2 & ~ & ~ \\ \hline
        SPT-SJ221943-5254.4 & 334.93143 & -52.907192 & 19.7 & 33.7 & ~ & ~ \\ \hline
        SPT-SJ221952-6333.5 & 334.9679 & -63.559307 & 27.3 & 50.9 & ~ & ~ \\ \hline
        SPT-SJ222658-6237.8 & 336.74191 & -62.630619 & 10.3 & 20.2 & 2.38 & 45.4 \\ \hline
        SPT-SJ222707-5219.9 & 336.78314 & -52.332886 & 36.5 & 55.4 & ~ & ~ \\ \hline
        SPT-SJ223011-6310.7 & 337.54657 & -63.178612 & 24.6 & 45.8 & ~ & ~ \\ \hline
        SPT-SJ223042-5415.2 & 337.67706 & -54.254856 & 20.4 & 30.2 & ~ & ~ \\ \hline
        SPT-SJ223044-6329.4 & 337.68417 & -63.490105 & 16.1 & 29.1 & ~ & ~ \\ \hline
        SPT-SJ223108-6231.3 & 337.7861 & -62.522224 & 106.0 & 167.0 & ~ & ~ \\ \hline
        SPT-SJ223422-5405.9 & 338.59454 & -54.099758 & 65.2 & 94.5 & 1.98 & 45.9 \\ \hline
        SPT-SJ223825-5114.3 & 339.60501 & -51.239658 & 51.3 & 60.4 & 0.569 & 44.2 \\ \hline
        SPT-SJ223905-5526.0 & 339.77454 & -55.434715 & 58.3 & 87.4 & 0.482 & 44.1 \\ \hline
        SPT-SJ223911-5701.0 & 339.79987 & -57.016701 & 300.0 & 480.0 & 0.282 & 44.2 \\ \hline
        SPT-SJ224017-5232.0 & 340.07443 & -52.533623 & 34.5 & 53.4 & 0.392 & 43.6 \\ \hline
        SPT-SJ224223-6044.3 & 340.5986 & -60.739792 & 52.2 & 86.3 & ~ & ~ \\ \hline
        SPT-SJ224306-6250.9 & 340.77875 & -62.849735 & 207.0 & 317.0 & 1.0 & 45.5 \\ \hline
        SPT-SJ224353-5656.3 & 340.97388 & -56.938587 & 33.3 & 57.9 & 1.32 & 45.1 \\ \hline
        SPT-SJ224500-4931.8 & 341.25177 & -49.530224 & 19.3 & 35.1 & ~ & ~ \\ \hline
        SPT-SJ224615-5607.7 & 341.56628 & -56.128754 & 151.0 & 233.0 & ~ & ~ \\ \hline
        SPT-SJ224836-5933.9 & 342.15182 & -59.565441 & 27.5 & 45.3 & ~ & ~ \\ \hline
        SPT-SJ225136-5453.0 & 342.90167 & -54.884426 & 32.3 & 52.1 & ~ & ~ \\ \hline
        SPT-SJ225139-5801.0 & 342.91406 & -58.017036 & 14.1 & 23.3 & ~ & ~ \\ \hline
        SPT-SJ225400-5025.1 & 343.50137 & -50.419308 & 13.4 & 18.9 & 1.92 & 45.2 \\ \hline
        SPT-SJ225423-5148.7 & 343.59897 & -51.811817 & 54.0 & 71.0 & 0.55 & 44.2 \\ \hline
        SPT-SJ225457-5925.9 & 343.74026 & -59.431831 & 125.0 & 198.0 & 0.43 & 44.3 \\ \hline
        SPT-SJ225550-5245.7 & 343.96121 & -52.762154 & 31.3 & 55.5 & 0.15 & 42.5 \\ \hline
        SPT-SJ225647-5158.7 & 344.19833 & -51.978554 & 14.3 & 23.4 & ~ & ~ \\ \hline
        SPT-SJ230137-5913.2 & 345.40421 & -59.221107 & 57.7 & 87.9 & ~ & ~ \\ \hline
        SPT-SJ230352-5406.6 & 345.9693 & -54.110813 & 106.0 & 157.0 & ~ & ~ \\ \hline
        SPT-SJ230805-5218.9 & 347.02148 & -52.316448 & 43.0 & 60.2 & 0.153 & 42.7 \\ \hline
        SPT-SJ230903-5050.9 & 347.26373 & -50.849403 & 86.2 & 115.0 & ~ & ~ \\ \hline
        SPT-SJ231028-5941.1 & 347.62054 & -59.68594 & 13.7 & 36.0 & ~ & ~ \\ \hline
        SPT-SJ231102-6045.2 & 347.75952 & -60.753582 & 26.0 & 42.3 & 0.262 & 43.0 \\ \hline
        SPT-SJ231109-5624.7 & 347.78952 & -56.412251 & 12.1 & 20.5 & 0.808 & 44.0 \\ \hline
        SPT-SJ231327-5343.4 & 348.36429 & -53.723717 & 20.0 & 31.9 & 1.76 & 45.3 \\ \hline
        SPT-SJ231545-5018.6 & 348.93875 & -50.310604 & 516.0 & 700.0 & ~ & ~ \\ \hline
        SPT-SJ231653-5507.8 & 349.22244 & -55.131046 & 21.9 & 33.1 & 0.0958 & 42.0 \\ \hline
        SPT-SJ231716-6215.3 & 349.3194 & -62.256336 & 48.8 & 76.6 & 0.588 & 44.2 \\ \hline
        SPT-SJ231916-5331.8 & 349.81848 & -53.530834 & 33.7 & 52.3 & 1.35 & 45.1 \\ \hline
        SPT-SJ232047-5309.7 & 350.19983 & -53.161743 & 37.4 & 47.8 & ~ & ~ \\ \hline
        SPT-SJ232100-5402.5 & 350.25061 & -54.042355 & 38.7 & 59.9 & 1.42 & 45.3 \\ \hline
        SPT-SJ232142-6438.1 & 350.42841 & -64.636238 & 41.1 & 57.3 & ~ & ~ \\ \hline
        SPT-SJ232252-5322.0 & 350.7189 & -53.366703 & 36.4 & 48.4 & 0.181 & 42.8 \\ \hline
        SPT-SJ232510-5656.4 & 351.29572 & -56.940929 & 14.0 & 23.3 & 0.518 & 43.6 \\ \hline
        SPT-SJ232804-5630.0 & 352.02008 & -56.500526 & 20.4 & 33.1 & 1.03 & 44.6 \\ \hline
        SPT-SJ232920-4955.6 & 352.33704 & -49.927555 & 614.0 & 795.0 & 1.23 & 46.3 \\ \hline
        SPT-SJ233444-5251.2 & 353.68658 & -52.854595 & 185.0 & 306.0 & 1.17 & 45.7 \\ \hline
        SPT-SJ233612-5236.3 & 354.05023 & -52.605278 & 283.0 & 450.0 & ~ & ~ \\ \hline
        SPT-SJ233723-5216.2 & 354.34653 & -52.27103 & 30.9 & 56.2 & 1.35 & 45.1 \\ \hline
        SPT-SJ233726-5901.1 & 354.36047 & -59.018414 & 45.0 & 68.6 & ~ & ~ \\ \hline
        SPT-SJ233913-5523.7 & 354.80457 & -55.396275 & 12.5 & 21.4 & ~ & ~ \\ \hline
        SPT-SJ234118-5816.0 & 355.32788 & -58.267105 & 19.6 & 41.0 & 1.24 & 44.8 \\ \hline
        SPT-SJ234316-5709.4 & 355.82077 & -57.158314 & 38.2 & 56.2 & ~ & ~ \\ \hline
        SPT-SJ234327-5626.2 & 355.86276 & -56.437298 & 102.0 & 144.0 & 1.75 & 46.0 \\ \hline
        SPT-SJ234442-6108.2 & 356.17517 & -61.137993 & 25.2 & 46.3 & 0.643 & 44.1 \\ \hline
        SPT-SJ234720-5110.6 & 356.83377 & -51.176849 & 224.0 & 306.0 & ~ & ~ \\ \hline
        SPT-SJ234743-4946.4 & 356.93262 & -49.774513 & 153.0 & 217.0 & ~ & ~ \\ \hline
        SPT-SJ234751-5915.5 & 356.96622 & -59.259663 & 33.6 & 54.6 & 1.58 & 45.4 \\ \hline
        SPT-SJ234825-6049.2 & 357.10461 & -60.821117 & 98.8 & 145.0 & ~ & ~ \\ \hline
        SPT-SJ234845-5847.0 & 357.18857 & -58.784187 & 14.2 & 24.0 & ~ & ~ \\ \hline
        SPT-SJ234925-4932.4 & 357.35828 & -49.540848 & 220.0 & 282.0 & 0.413 & 44.5 \\ \hline
        SPT-SJ235322-5857.3 & 358.34457 & -58.955879 & 19.3 & 35.6 & 1.01 & 44.5 \\ \hline
        SPT-SJ235540-5418.4 & 358.91727 & -54.30719 & 16.4 & 21.2 & ~ & ~ \\ \hline
        SPT-SJ235753-5311.2 & 359.47214 & -53.187664 & 830.0 & 1220.0 & ~ & ~ \\ \hline
        SPT-SJ235844-6053.0 & 359.68713 & -60.884502 & 142.0 & 270.0 & ~ & ~ \\ \hline
        SPT-SJ235903-6055.1 & 359.76535 & -60.919224 & 52.9 & 110.0 & 0.0962 & 42.3 \\ \hline
        SPT-SJ235920-6057.0 & 359.83713 & -60.950008 & 68.7 & 148.0 & ~ & ~ \\ \hline
        SPT-SJ235947-5042.5 & 359.94876 & -50.709415 & 47.0 & 47.7 & ~ & ~  

\label{table:sources}
\end{longtable}
\end{center}

\end{document}

%% file: authors.tex
\shortauthors{J.~C.~Hood, et al.}
\author[0000-0003-4157-4185]{J.~C.~Hood II}
\affiliation{Kavli Institute for Cosmological Physics, University of Chicago, 5640 South Ellis Avenue, Chicago, IL, USA 60637}
\affiliation{Department of Astronomy and Astrophysics, University of Chicago, 5640 South Ellis Avenue, Chicago, IL, USA 60637}
\author{P.~A.~R.~Ade} 
\affiliation{Cardiff University, Cardiff CF10 3XQ, United Kingdom}
\author[0000-0002-4435-4623]{A.~J.~Anderson}
\affiliation{Fermi National Accelerator Laboratory, MS209, P.O. Box 500, Batavia, IL 60510}
\author[0000-0002-0517-9842]{M.~Archipley}\affiliation{Department of Astronomy and Astrophysics, University of Chicago, 5640 South Ellis Avenue, Chicago, IL, USA 60637}
\affiliation{Kavli Institute for Cosmological Physics, University of Chicago, 5640 South Ellis Avenue, Chicago, IL, USA 60637}
\author{J.~E.~Austermann} \affiliation{NIST Quantum Devices Group, 325 Broadway Mailcode 817.03, Boulder, CO, USA 80305} \affiliation{Department of Physics, University of Colorado, Boulder, CO, USA 80309}
\author{J.~A.~Beall} \affiliation{NIST Quantum Devices Group, 325 Broadway Mailcode 817.03, Boulder, CO, USA 80305}
\author[0000-0001-5868-0748]{A.~N.~Bender} \affiliation{High Energy Physics Division, Argonne National Laboratory, 9700 S. Cass Avenue, Argonne, IL, USA 60439} \affiliation{Kavli Institute for Cosmological Physics, University of Chicago, 5640 South Ellis Avenue, Chicago, IL, USA 60637}
\author[0000-0002-5108-6823]{B.~A.~Benson} \affiliation{Fermi National Accelerator Laboratory, MS209, P.O. Box 500, Batavia, IL 60510} \affiliation{Kavli Institute for Cosmological Physics, University of Chicago, 5640 South Ellis Avenue, Chicago, IL, USA 60637} \affiliation{Department of Astronomy and Astrophysics, University of Chicago, 5640 South Ellis Avenue, Chicago, IL, USA 60637}
\author[0000-0003-4847-3483]{F.~Bianchini} \affiliation{School of Physics, University of Melbourne, Parkville, VIC 3010, Australia}
\author[0000-0001-7665-5079]{L.~E.~Bleem} \affiliation{High Energy Physics Division, Argonne National Laboratory, 9700 S. Cass Avenue, Argonne, IL, USA 60439} \affiliation{Kavli Institute for Cosmological Physics, University of Chicago, 5640 South Ellis Avenue, Chicago, IL, USA 60637}\affiliation{Department of Astronomy and Astrophysics, University of Chicago, 5640 South Ellis Avenue, Chicago, IL, USA 60637}
\author[0000-0002-2044-7665]{J.~E.~Carlstrom} \affiliation{Kavli Institute for Cosmological Physics, University of Chicago, 5640 South Ellis Avenue, Chicago, IL, USA 60637} \affiliation{Department of Physics, University of Chicago, 5640 South Ellis Avenue, Chicago, IL, USA 60637} \affiliation{High Energy Physics Division, Argonne National Laboratory, 9700 S. Cass Avenue, Argonne, IL, USA 60439} \affiliation{Department of Astronomy and Astrophysics, University of Chicago, 5640 South Ellis Avenue, Chicago, IL, USA 60637} \affiliation{Enrico Fermi Institute, University of Chicago, 5640 South Ellis Avenue, Chicago, IL, USA 60637}
\author{C.~L.~Chang} \affiliation{Kavli Institute for Cosmological Physics, University of Chicago, 5640 South Ellis Avenue, Chicago, IL, USA 60637} \affiliation{High Energy Physics Division, Argonne National Laboratory, 9700 S. Cass Avenue, Argonne, IL, USA 60439} \affiliation{Department of Astronomy and Astrophysics, University of Chicago, 5640 South Ellis Avenue, Chicago, IL, USA 60637}
\author{P.~Chaubal} \affiliation{School of Physics, University of Melbourne, Parkville, VIC 3010, Australia}
\author{H.~C.~Chiang} \affiliation{Department of Physics, McGill University, 3600 Rue University, Montreal, Quebec H3A 2T8, Canada} \affiliation{School of Mathematics, Statistics \& Computer Science, University of KwaZulu-Natal, Durban, South Africa}
\author[0000-0002-3091-8790]{T-L.~Chou} \affiliation{Kavli Institute for Cosmological Physics, University of Chicago, 5640 South Ellis Avenue, Chicago, IL, USA 60637} \affiliation{Department of Physics, University of Chicago, 5640 South Ellis Avenue, Chicago, IL, USA 60637}
\author{R.~Citron} \affiliation{University of Chicago, 5640 South Ellis Avenue, Chicago, IL, USA 60637}
\author{C.~Corbett~Moran} \affiliation{TAPIR, Walter Burke Institute for Theoretical Physics, California Institute of Technology, 1200 E California Blvd, Pasadena, CA, USA 91125}
\author[0000-0001-9000-5013]{T.~M.~Crawford} \affiliation{Kavli Institute for Cosmological Physics, University of Chicago, 5640 South Ellis Avenue, Chicago, IL, USA 60637} \affiliation{Department of Astronomy and Astrophysics, University of Chicago, 5640 South Ellis Avenue, Chicago, IL, USA 60637}
\author{A.~T.~Crites} \affiliation{Kavli Institute for Cosmological Physics, University of Chicago, 5640 South Ellis Avenue, Chicago, IL, USA 60637} \affiliation{Department of Astronomy and Astrophysics, University of Chicago, 5640 South Ellis Avenue, Chicago, IL, USA 60637} \affiliation{Dunlap Institute for Astronomy \& Astrophysics, University of Toronto, 50 St George St, Toronto, ON, M5S 3H4, Canada} \affiliation{Department of Astronomy \& Astrophysics, University of Toronto, 50 St George St, Toronto, ON, M5S 3H4, Canada}
\author[0000-0001-5105-9473]{T.~de~Haan} \affiliation{Institute of Particle and Nuclear Studies (IPNS), High Energy Accelerator Research Organization (KEK), Tsukuba, Ibaraki 305-0801, Japan} \affiliation{International Center for Quantum-field Measurement Systems for Studies of the Universe and Particles (QUP), High Energy Accelerator Research Organization (KEK), Tsukuba, Ibaraki 305-0801, Japan}
\author{M.~A.~Dobbs} \affiliation{Department of Physics, McGill University, 3600 Rue University, Montreal, Quebec H3A 2T8, Canada} \affiliation{Canadian Institute for Advanced Research, CIFAR Program in Gravity and the Extreme Universe, Toronto, ON, M5G 1Z8, Canada}
\author{W.~Everett} \affiliation{Department of Astrophysical and Planetary Sciences, University of Colorado, Boulder, CO, USA 80309}
\author[0000-0002-7145-1824]{A.~Foster}\affiliation{Joseph Henry Laboratories of Physics, Jadwin Hall, Princeton University, Princeton, NJ 08544, USA}
\author{J.~Gallicchio} \affiliation{Kavli Institute for Cosmological Physics, University of Chicago, 5640 South Ellis Avenue, Chicago, IL, USA 60637} \affiliation{Harvey Mudd College, 301 Platt Blvd., Claremont, CA 91711}
\author{E.~M.~George} \affiliation{European Southern Observatory, Karl-Schwarzschild-Str. 2, 85748 Garching bei M\"{u}nchen, Germany} \affiliation{Department of Physics, University of California, Berkeley, CA, USA 94720}
\author[0000-0001-7652-9451]{N.~Gupta} \affiliation{CSIRO Space \& Astronomy, PO Box 1130, Bentley WA 6102, Australia}
\author{N.~W.~Halverson} \affiliation{Department of Astrophysical and Planetary Sciences, University of Colorado, Boulder, CO, USA 80309} \affiliation{Department of Physics, University of Colorado, Boulder, CO, USA 80309}
\author{G.~C.~Hilton} \affiliation{NIST Quantum Devices Group, 325 Broadway Mailcode 817.03, Boulder, CO, USA 80305}
\author[0000-0002-0463-6394]{G.~P.~Holder} \affiliation{Astronomy Department, University of Illinois at Urbana-Champaign, 1002 W. Green Street, Urbana, IL 61801, USA} \affiliation{Department of Physics, University of Illinois Urbana-Champaign, 1110 W. Green Street, Urbana, IL 61801, USA} \affiliation{Canadian Institute for Advanced Research, CIFAR Program in Gravity and the Extreme Universe, Toronto, ON, M5G 1Z8, Canada}
\author{W.~L.~Holzapfel} \affiliation{Department of Physics, University of California, Berkeley, CA, USA 94720}
\author{J.~D.~Hrubes} \affiliation{University of Chicago, 5640 South Ellis Avenue, Chicago, IL, USA 60637}
\author[0000-0003-3595-0359]{N.~Huang} \affiliation{Department of Physics, University of California, Berkeley, CA, USA 94720}
\author{J.~Hubmayr} \affiliation{NIST Quantum Devices Group, 325 Broadway Mailcode 817.03, Boulder, CO, USA 80305}
\author{K.~D.~Irwin} \affiliation{SLAC National Accelerator Laboratory, 2575 Sand Hill Road, Menlo Park, CA 94025} \affiliation{Dept. of Physics, Stanford University, 382 Via Pueblo Mall, Stanford, CA 94305}
\author[0000-0001-9194-7168]{E.~J\"arvel\"a} \affiliation{Department of Physics \& Astronomy, Texas Tech University, Box 41051, Lubbock TX, 79409-1051}
\author{L.~Knox} \affiliation{Department of Physics, University of California, One Shields Avenue, Davis, CA, USA 95616}
\author{A.~T.~Lee} \affiliation{Department of Physics, University of California, Berkeley, CA, USA 94720} \affiliation{Physics Division, Lawrence Berkeley National Laboratory, Berkeley, CA, USA 94720}
\author{D.~Li} \affiliation{NIST Quantum Devices Group, 325 Broadway Mailcode 817.03, Boulder, CO, USA 80305} \affiliation{SLAC National Accelerator Laboratory, 2575 Sand Hill Road, Menlo Park, CA 94025}
\author[0000-0002-4747-4276]{A.~Lowitz} \affiliation{Department of Astronomy and Astrophysics, University of Chicago, 5640 South Ellis Avenue, Chicago, IL, USA 60637}
\author[0000-0003-0976-4755]{T.~J.~Maccarone} \affiliation{Department of Physics \& Astronomy, Texas Tech University, Box 41051, Lubbock TX, 79409-1051}
\author{J.~J.~McMahon} \affiliation{Kavli Institute for Cosmological Physics, University of Chicago, 5640 South Ellis Avenue, Chicago, IL, USA 60637} \affiliation{Department of Physics, University of Chicago, 5640 South Ellis Avenue, Chicago, IL, USA 60637} \affiliation{Department of Astronomy and Astrophysics, University of Chicago, 5640 South Ellis Avenue, Chicago, IL, USA 60637}
\author{S.~S.~Meyer} \affiliation{Kavli Institute for Cosmological Physics, University of Chicago, 5640 South Ellis Avenue, Chicago, IL, USA 60637} \affiliation{Department of Physics, University of Chicago, 5640 South Ellis Avenue, Chicago, IL, USA 60637} \affiliation{Department of Astronomy and Astrophysics, University of Chicago, 5640 South Ellis Avenue, Chicago, IL, USA 60637} \affiliation{Enrico Fermi Institute, University of Chicago, 5640 South Ellis Avenue, Chicago, IL, USA 60637}
\author{J.~Montgomery} \affiliation{Department of Physics, McGill University, 3600 Rue University, Montreal, Quebec H3A 2T8, Canada}
\author{T.~Natoli} \affiliation{Department of Astronomy and Astrophysics, University of Chicago, 5640 South Ellis Avenue, Chicago, IL, USA 60637} \affiliation{Kavli Institute for Cosmological Physics, University of Chicago, 5640 South Ellis Avenue, Chicago, IL, USA 60637}
\author{J.~P.~Nibarger} \affiliation{NIST Quantum Devices Group, 325 Broadway Mailcode 817.03, Boulder, CO, USA 80305}
\author[0000-0002-5254-243X]{G.~Noble} \affiliation{Department of Physics, McGill University, 3600 Rue University, Montreal, Quebec H3A 2T8, Canada}
\author{V.~Novosad} \affiliation{Materials Sciences Division, Argonne National Laboratory, 9700 S. Cass Avenue, Argonne, IL, USA 60439}
\author{S.~Padin} \affiliation{Kavli Institute for Cosmological Physics, University of Chicago, 5640 South Ellis Avenue, Chicago, IL, USA 60637} \affiliation{Department of Astronomy and Astrophysics, University of Chicago, 5640 South Ellis Avenue, Chicago, IL, USA 60637} \affiliation{California Institute of Technology, 1200 E. California Blvd., Pasadena CA 91125 USA}
\author{S.~Patil} \affiliation{School of Physics, University of Melbourne, Parkville, VIC 3010, Australia}
\author[0000-0001-7946-557X]{K.~A.~Phadke}\affiliation{Astronomy Department, University of Illinois at Urbana-Champaign, 1002 W. Green Street, Urbana, IL 61801, USA} \affiliation{Department of Physics, University of Illinois Urbana-Champaign, 1110 W. Green Street, Urbana, IL 61801, USA} \affiliation{Center for AstroPhysical Surveys, National Center for Supercomputing Applications, Urbana, IL, 61801, USA}
\author{C.~Pryke} \affiliation{School of Physics and Astronomy, University of Minnesota, 116 Church Street S.E. Minneapolis, MN, USA 55455}
\author[0000-0003-2226-9169]{C.~L.~Reichardt} \affiliation{School of Physics, University of Melbourne, Parkville, VIC 3010, Australia}
\author{J.~E.~Ruhl} \affiliation{Department of Physics, Case Western Reserve University, Cleveland, OH, 44106, USA} \affiliation{Center for Education and Research in Cosmology and Astrophysics, Case Western Reserve University, Cleveland, OH, USA 44106}
\author{B.~R.~Saliwanchik} \affiliation{Physics Department, Center for Education and Research in Cosmology and Astrophysics, Case Western Reserve University, Cleveland, OH, USA 44106} \affiliation{Department of Physics, Yale University, P.O. Box 208120, New Haven, CT 06520-8120}
\author{K.~K.~Schaffer} \affiliation{Kavli Institute for Cosmological Physics, University of Chicago, 5640 South Ellis Avenue, Chicago, IL, USA 60637} \affiliation{Enrico Fermi Institute, University of Chicago, 5640 South Ellis Avenue, Chicago, IL, USA 60637} \affiliation{Liberal Arts Department, School of the Art Institute of Chicago, 112 S Michigan Ave, Chicago, IL, USA 60603}
\author{C.~Sievers} \affiliation{University of Chicago, 5640 South Ellis Avenue, Chicago, IL, USA 60637}
\author{A.~Simpson} \affiliation{Department of Astronomy and Astrophysics, University of Chicago, 5640 South Ellis Avenue, Chicago, IL, USA 60637}
\author{G.~Smecher} \affiliation{Department of Physics, McGill University, 3600 Rue University, Montreal, Quebec H3A 2T8, Canada} \affiliation{Three-Speed Logic, Inc., Victoria, B.C., V8S 3Z5, Canada}
\author{A.~A.~Stark} \affiliation{Harvard-Smithsonian Center for Astrophysics, 60 Garden Street, Cambridge, MA, USA 02138}
\author[0000-0002-2077-6004]{C.~Tandoi}\affiliation{Astronomy Department, University of Illinois at Urbana-Champaign, 1002 W. Green Street, Urbana, IL 61801, USA}
\author{C.~Tucker} \affiliation{Cardiff University, Cardiff CF10 3XQ, United Kingdom}
\author{T.~Veach} \affiliation{Space Science and Engineering Division, Southwest Research Institute, San Antonio, TX 78238}
\author[0000-0001-7192-3871]{J.~D.~Vieira} \affiliation{Astronomy Department, University of Illinois at Urbana-Champaign, 1002 W. Green Street, Urbana, IL 61801, USA} \affiliation{Department of Physics, University of Illinois Urbana-Champaign, 1110 W. Green Street, Urbana, IL 61801, USA} \affiliation{Center for AstroPhysical Surveys, National Center for Supercomputing Applications, Urbana, IL, 61801, USA}
\author{G.~Wang} \affiliation{High Energy Physics Division, Argonne National Laboratory, 9700 S. Cass Avenue, Argonne, IL, USA 60439}
\author[0000-0002-3157-0407]{N.~Whitehorn} \affiliation{Department of Physics and Astronomy, Michigan State University, 567 Wilson Road, East Lansing, MI 48824}
\author[0000-0001-5411-6920]{W.~L.~K.~Wu} \affiliation{Kavli Institute for Cosmological Physics, University of Chicago, 5640 South Ellis Avenue, Chicago, IL, USA 60637} \affiliation{SLAC National Accelerator Laboratory, 2575 Sand Hill Road, Menlo Park, CA 94025}
\author{V.~Yefremenko} \affiliation{High Energy Physics Division, Argonne National Laboratory, 9700 S. Cass Avenue, Argonne, IL, USA 60439}
\author{J.~A.~Zebrowski} \affiliation{Department of Physics, University of California, Berkeley, CA, USA 94720}